\documentclass[journal,onecolumn,12pt,draftclsnofoot]{IEEEtran}
\usepackage{graphicx,psfrag,epsfig,epsf,latexsym,hhline,amsmath,amssymb,multirow}
\usepackage[usenames,dvipsnames]{pstricks}
\usepackage{pst-plot}
\usepackage{subcaption}
\usepackage{hyperref}
\usepackage{graphicx}
\usepackage{amsthm}
\usepackage{algorithm}
\usepackage{algpseudocode}
\usepackage[noadjust]{cite}
\usepackage{blindtext}
\usepackage{etoolbox,tcolorbox}
\usepackage{stfloats}
\input{Jerry.def}

\begin{document}
\title{Iterative Collision Resolution for Slotted ALOHA with NOMA for Heterogeneous Devices}
\author{Yu-Chih Huang, Shin-Lin Shieh, Yu-Pin Hsu, and Hao-Ping Cheng

\thanks{Y.-C. Huang is with the Institute of Communications Engineering, National Chiao Tung University, Hsinchu, Taiwan (email: jerryhuang@nctu.edu.tw). S.-L. Shieh and Y.-P. Hsu are with the Department of Communication Engineering, National Taipei University, New Taipei City, Taiwan (email: \{yupinhsu, slshieh\}@mail.ntpu.edu.tw). H.-P. Cheng was with the Department of Communication Engineering, National Taipei University, New Taipei City, Taiwan and is now with the MediaTek Inc.}

}

\maketitle

\begin{abstract}
    In this paper, the problem of using uncoordinated multiple access (UMA) to serve a massive amount of heterogeneous users is investigated.
    Leveraging the heterogeneity, we propose a novel UMA protocol, called iterative collision resolution for slotted ALOHA (IRSA) with non-orthogonal multiple access (NOMA), to improve the conventional IRSA.
    In addition to the inter-slot successive interference cancellation (SIC) technique used in existing IRSA-based schemes, the proposed protocol further employs the intra-slot SIC technique that enables collision resolution for certain configurations of collided packets. A novel multi-dimensional density evolution is then proposed to analyze and to optimize the proposed protocol. Simulation results show that the proposed IRSA with NOMA protocol can efficiently exploit the heterogeneity among users and the multi-dimensional density evolution can accurately predict the throughput performance. Last, \textcolor{black}{an extension of the proposed IRSA with NOMA protocol to the frame-asynchronous setting is investigated}, where a boundary effect similar to that in spatially-coupled low-density parity check codes can be observed to bootstrap the decoding process.
\end{abstract}


%------------------------------------------------------------------------

\section{Introduction}\label{sec:intro}
To realize the overarching ambition of internet of things (IoT) \cite{IOT_white11, IOT_survey15}, the current 3GPP Specification of the fifth generation cellular network technology has defined a use scenario, termed massive machine-type communication (mMTC) \cite{TR22.891, mmtc_survey18}. This use scenario is expected to accommodate a massive number of users with sporadic activities. Moreover, in such applications, the payload is assumed to be rather small. In this circumstance, traditional philosophy of first coordinating then communicating becomes extremely inefficient, as there are so many users to coordinate while so small amount of data for each user to transmit. Hence, uncoordinated multiple access (UMA) techniques that can communicate without first establishing coordination are highly desirable.

Among many UMA techniques, slotted ALOHA (SA) \cite{BertsekasGallager92} has been adopted in many practical communication systems. In an SA protocol, each user transmits a packet at the time slot next to that in which the data packet arrived. When multiple users send packets at the same slot, i.e., the packets collide, these collided packets are discarded \textcolor{black}{ (i.e., collision channel model)} and retransmissions are scheduled according to some back-off mechanism. Recently, this concept has even been adopted in modern cellular systems such as SigFoX and LoRaWAN \cite{Liva2019:white_paper}. Notwithstanding the success and popularity, it is well known that the conventional SA achieves an efficiency (to be defined in Section~\ref{sec:problem}) of at most $1/e\approx 36.8 \%$ \cite[Sec. 4.2]{BertsekasGallager92}.
%In contrast, when ignoring the resource used for establishing coordination, a perfectly coordinated multiple access scheme can achieve the efficiency of $100\%$.
To improve the efficiency, Casini {\it et al.} in \cite{Casini:CRDSA} proposed contention resolution diversity slotted ALOHA (CRDSA) that improves the efficiency to be at most $55\%$ \textcolor{black}{asymptotically as the frame size goes to infinity}. In CRDSA, each active user (those having data to send) sends replicated packets on two\footnote{\textcolor{black}{Higher efficiencies can be achieved by sending more replicas.}} randomly selected slots . The collided packets are not discarded in CRDSA; instead, they are stored with the hope that inter-slot\footnote{Here, we particularly highlight the term ``inter-slot" for distinguishing it from the intra-slot SIC discussed in Section~\ref{sec:problem}.} successive interference cancellation (SIC) can resolve collisions by their replicas. Due to its improved efficiency, a version of CRDSA has been included into the digital video broadcasting (DVB) standardization \cite{DVB14}.

After the paradigm-shifting idea \cite{Casini:CRDSA}, much effort has been devoted to improve CRDSA by making connections to error-correction codes \cite{Liva2011:IRSA_TCOM}. To the best of our knowledge, the connection was first made in \cite[Example 2]{Liva2011:IRSA_TCOM} where a CRDSA system was shown to have a left-regular bipartite graph representation and inter-slot SIC is interpreted as belief propagation (BP) decoding \cite{urbanke_book}. After this connection, Liva in \cite{Liva2011:IRSA_TCOM} then proposed irregular repetition slotted ALOHA (IRSA), in which each active user randomly chooses the number of repetitions and randomly selects time slots to send those replicas. It can be observed that an IRSA scheme corresponds to an irregular bipartite graph. It was shown in \cite{Liva2011:IRSA_TCOM} that by optimizing the left degree distribution with the help of density evolution \cite{urbanke_book}, one can achieve a significantly improved efficiency of $96.5\%$ \textcolor{black}{asymptotically}. It was then recognized in \cite{Krishna12_IRSA} that the coding problem induced by IRSA is similar to that of LT codes \cite{LT_code02} with the peeling decoder \cite{Luby01peeling}. Remarkably, with this observation, Narayanan and Pfister showed that applying the robust Soliton distribution as the left degree distribution of IRSA achieves the efficiency of $100\%$ asymptotically \cite{Krishna12_IRSA}, which is optimal among all IRSA schemes. \textcolor{black}{Also, in \cite{Liva12spatially_couple}, Liva {\it et al.} proposed grouping many frames together to form a super frame and employing spatially-coupled IRSA on the super frame to achieve the efficiency of $100\%$ asymptotically.} The performance of these UMA schemes were then assessed in a more realistic model in \cite{stefanovi18}.

Several variants of IRSA have also been proposed and analyzed. In \cite{iAmat17}, the frame-asynchronous version of IRSA was investigated and a boundary effect similar to that exhibited in spatially-coupled low-density parity-check (LDPC) codes \cite{urbanke11_SC_LDPC} is observed. Thanks to this boundary effect, it was shown that regular left degree distributions suffice to attain the optimal efficiency for frame-asynchronous IRSA. In \cite{Liva2015:CodedAloha}, a coded slotted ALOHA protocol was proposed where maximum distance separable (MDS) codes replace repetition codes when sending replicas. It was shown that this class of schemes can achieve rates higher than that obtained by IRSA. In \cite{Jinhong19}, a combination of IRSA and physical-layer network coding \cite{zhang06} (a.k.a. compute-and-forward \cite{nazer2011CF}) was proposed, where the receiver decodes the received signal at a slot to an integer linear combination of collided packets. These combinations are later used to recover individual packets that would not be decodable in conventional IRSA. In contrast to the aforementioned works focusing on the asymptotic analysis, another series of work in the literature investigated non-asymptotic analysis of IRSA or coded slotted ALOHA. For example, using packet loss rate and frame error rate as the metrics, the performance of IRSA in the waterfall region was analyzed in \cite{iAmat18}. Moreover, the performance of coded slotted ALOHA in the error floor region and that in the waterfall region were separately investigated in \cite{iAmat15_error_floor} and \cite{Fereydounian19}, respectively.

{\color{black} Another highly related line of research named unsourced multiple access was initiated by Polyanskiy in \cite{polyanskiy17uma}, where a large number of homogeneous users wish to communicate with a receiver who is only interested in recovering as many users' data as possible without concern for their identities. Using random Gaussian codebooks together with maximum likelihood decoding, \cite{polyanskiy17uma} presented finite block-length achievability bounds. A low-complexity coding scheme which employs the concatenation of an inner compute-and-forward code \cite{nazer2011CF} and a outer binary adder channel code was then introduced and analyzed for unsourced multiple access in \cite{ordentlich17uma}. The current state of the art is \cite{krishna20uma} where a coded compress sensing algorithm followed by a tree-like code was proposed and analyzed. }

%However, computing integer linear combinations according to real linear combination (induced by the channel) usually involves large encoding and decoding complexity of coding over prime fields. Moreover, even with some recent progresses \cite{Jinhong13,Engin15, Huang18_CF}, the mismatch between the integer and real linear combinations still causes significant reduction in achievable rate.

Most of the works in the UMA literature, including those discussed above, considered homogeneous users. However, in many applications in the mMTC use scenarios, several heterogeneous types of users coexist \cite{mmtc_survey18, durisi18, TR36.866}. Therefore, it is imperative to understand how the heterogeneity would influence the system performance and it is crucial to design UMA schemes that can exploit such heterogeneity among users.

To address the heterogeneity among users in UMA, in this work, we incorporate the power-domain non-orthogonal multiple access (NOMA) \cite{mazzini98, Dai15, Saito13VtcS, Saito13Pimrc} into IRSA and proposed the IRSA with NOMA protocol. In the proposed scheme, leveraging the power-domain NOMA in the physical layer, when packets coming from heterogenous types of users collide at a slot, it is possible that all the packets can be resolved by intra-slot SIC. These decoded packets are then used to resolve collisions in other slots by inter-slot SIC. We must emphasize that this is in sharp contrast to the idea of multi-packet reception \cite{verdu_MPR88}. In SA with multi-packet reception \cite{Ghanbarinejad13, stefanovi18cl}, as long as the number of collided packets at a slot is smaller than some predefined parameter, all the packets can be decoded. Evidently, those users are still homogeneous and the benefit of having heterogeneous users remains unexploited. However, in our proposed IRSA with NOMA, not only the number of collided packets but also the types of users who transmit the packets will affect the decodability\footnote{A more detailed comparison between SA with multi-packet reception and that with NOMA is relegated to Section~\ref{sec:problem}.}.

The main contributions of this paper are provided as follows.
\begin{itemize}
  \item A novel IRSA protocol that incorporates the power-domain NOMA, namely the IRSA with NOMA scheme, is proposed. In such a protocol, depending on the types of users, multiple collided packets at a time slot may be decodable by intra-slot SIC. The decoded packets are then used to resolve collisions in other slots by inter-slot SIC.
  \item The connection between the proposed IRSA with NOMA and a bipartite graph with heterogeneous variable nodes and special check nodes is drawn. This connection is then leveraged to propose a novel multi-dimensional density evolution, which assists us in analyzing the proposed IRSA with NOMA protocol.
  \item According to the proposed multi-dimensional density evolution, a constrained optimization problem for maximizing the efficiency is formulated. Through solving this constrained optimization problem, the best degree distribution of each type of users for the proposed IRSA with NOMA protocol can be found.
  \item Extensive simulations are provided. For the considered simulation environment with a practical frame size, it is shown that when there are two and three heterogeneous types, the proposed scheme achieves the sum (over types) efficiencies of up to $132\%$ ($143\%$ asymptotically) and $170\%$ ($185\%$ asymptotically), respectively. Moreover, it also indicates that the analysis based on the proposed multi-dimensional density evolution can accurately predict the actual performance of the proposed scheme.
  \item The extension of the proposed protocol to the frame-asynchronous IRSA is presented and the corresponding multi-dimensional density evolution is analyzed. It is observed that under the frame-asynchronous circumstance, the proposed IRSA with NOMA also exhibits a boundary effect similar to that in \cite{iAmat17}. Hence, left-regular degree distributions again suffice to achieve performance that is as good as that achieved by an optimal degree distribution.
  %\item Last but not least, we revisit the problem of UMA with homogeneous users. As an extension, we enable the proposed NOMA with IRSA protocol for this homogeneous setting by allowing each user to randomly select a power level for sending its packet. %Moreover, we discuss a use scenario of this extension in low earth orbit (LEO) communication systems.
\end{itemize}

\textcolor{black}{During the revision of this paper, we became aware of a pioneering work \cite{Gaudenzi17power_unbalance}, in which CRDSA with power unbalance is investigated from a joint physical and MAC layers perspective. By allowing each user randomly selects its power according to a log-normal distribution, Mengali \textit{et al.} show that significant gains can be obtained by intra-slot SIC. However, how to exploit heterogeneity inherent in the network, how to analyze the asymptotic performance, and how to obtain optimal degree distributions when there are multiple types in the network are left undiscussed.}

\subsection{Organization}
In Section~\ref{sec:problem}, we formally state the problem of UMA with heterogeneous types of users and propose the IRSA with NOMA protocol. In Section~\ref{sec:graph}, we present a graph representation of the proposed protocol and translate the decoding process as a modified peeling decoder for the corresponding graph. In Section~\ref{sec:DE}, to analyze the asymptotic performance of the proposed protocol, we propose a multi-dimensional density evolution technique, with which we formulate an optimization problem for obtaining optimal policies for our proposed protocol. Optimal policies are then obtained by the differential evolution technique. Simulation results are presented in Section~\ref{sec:simulation}, followed by \textcolor{black}{the extension of the proposed IRSA with NOMA scheme to the frame asynchronous setting} in Sections~\ref{sec:FA}. Finally, some conclusion remarks are given in Section~\ref{sec:conclude}.

\subsection{Notational Convention}
We use $\mc{F}[x]$ to denote a polynomial with variable $x$ and we denote by $\mc{F}'[x]=\mathrm{d}\mc{F}[x]/\mathrm{d}x$ and $\mc{F}^{\{k\}}[x]=\mathrm{d^k\mc{F}[x]}/\mathrm{d} x^k$ the first and $k$-th derivative of $\mc{F}[x]$, respectively. We denote by $\mbb{N}$ the set of natural numbers. For two natural numbers $N_1, N_2\in\mbb{N}$ with $N_1 < N_2$, we denote by $[N_1]=\{1, 2, \ldots, N_1\}$ and $[N_1:N_2]=\{N_1, N_1+1, \ldots, N_2\}$.

\section{Problem Statement and Proposed IRSA with NOMA}\label{sec:problem}
\begin{figure}
    \centering
    \includegraphics[width=3.5in]{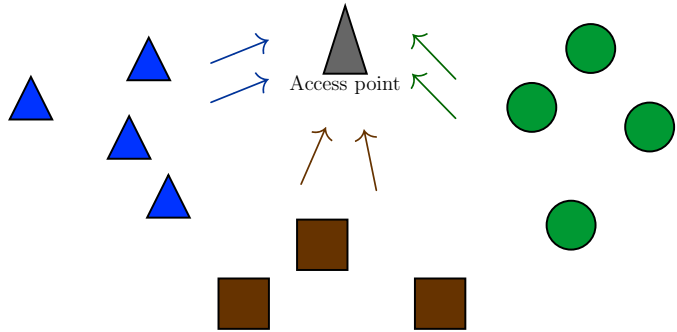}
    \caption{UMA with $T=3$ heterogeneous types of users.}
    \label{fig:mMTC}
\end{figure}

We consider an UMA setting where a massive number of users coming from $T\in\mbb{N}$ different types of applications wish to communicate with the base station as shown in Fig.~\ref{fig:mMTC}. Suppose time is slotted and synchronized. Furthermore, every $N\in\mbb{N}$ successive slots are organized into a frame; frames are also perfectly synchronous among users, as in other conventional IRSA schemes \cite{Casini:CRDSA, Liva2011:IRSA_TCOM, Krishna12_IRSA}. A frame-asynchronous version of this problem is relegated to Section~\ref{sec:FA}. A user who has data to send within a frame is said to be active in this frame. Potentially, there could be an arbitrary number of users for each type; Suppose $K^{(t)}\in\mbb{N}$ of type $t\in[T]$ users are active. Each active user of type $t\in[T]$ constructs $L$ replicas of its packet, where the value $L$ is sampled from a degree distribution $\mc{L}^{(t)}$; then, the user sends the $L$ replicas in $L$ slots out of $N$ slots chosen uniformly at random in the frame.  The degree distribution is our design issue for maximizing efficiency (that will be discussed soon).

\textcolor{black}{Leveraging the natural heterogeneity and the maturity of SIC technique \cite{SIC_chip18}, we propose the decoder that employs power-domain NOMA \cite{mazzini98, Dai15, Saito13Pimrc, Saito13VtcS} at each slot. Specifically, without loss of generality, we assume the decoding order of power-domain NOMA in the physical layer is given by $1\rightarrow 2\rightarrow\ldots\rightarrow T$ and we define the intra-SIC decodable patterns as follows.
\begin{define}(Type $t$ decodable pattern)
For a slot $n\in[N]$, we define the transmission pattern $\mathbf{c}_n=(c_1,\ldots,c_T)$ to be a $T$-dimensional vector where the entry $c_t$ represents the total number of unresolved type $t$ packets at this slot. Such a vector $\mathbf{c}_n$ is type $t$ decodable if and only if the following conditions hold:
\begin{align*}
    &\text{(i) $c_t=1$},\quad\quad\quad \text{(ii) $c_{t'}\leq 1$ for $t'\in[t-1]$}\\
    &\text{(iii) $c_{\tilde{t}}\leq \tilde{t}-t$ for $\tilde{t}\in[t+1:T]$,} \quad\text{(iv) $\sum_{\tilde{t}=t+1}^Tc_{\tilde{t}}\leq T-t$.}
\end{align*}
%\begin{itemize}
%    \item[(i)] $c_t=1$;
%    \item[(ii)] $c_{t'}\leq 1$ for $t'\in[t-1]$;
%    \item[(iii)] $c_{\tilde{t}}\leq \tilde{t}-t$ for $\tilde{t}\in[t+1:T]$;
%    \item[(iv)] $\sum_{\tilde{t}=t+1}^Tc_{\tilde{t}}\leq T-t$.
%\end{itemize}
\end{define}
The type $t$ decodable pattern is defined based on the principle of power-domain NOMA. In power-domain NOMA, it is assumed that a packet can be successfully decoded by intra-slot SIC as long as each interfered type has at most one packet collided at the same slot. This justifies the above conditions (i) and (ii). Moreover, types having later decoding order means that they will cause smaller interference. Therefore, \textit{trading} a type $\tilde{t}\in[t+1:T]$ packet for a type $\hat{t}>\tilde{t}$ packet would not affect the decodability, justifying the conditions (iii) and (iv). %This assumption will be justified in Remark~\ref{rmk:NOMAvsMPR}, in which comparison with multi-packet reception is also made.
\begin{define}(Type $t$ decodable set)
    The type $t$ decodable set $\mc{D}\upt$ consists of all the type $t$ decodable patterns. i.e., $\mc{D}\upt=\{\mathbf{c}:\mathbf{c}~\text{is type $t$ decodable}\}$.
\end{define}
\begin{example}\label{ex:decodable_set}
    When $T=2$, it is easy to check that $\mc{D}^{(1)}=\{(1,0),(1,1)\}$ and $\mc{D}^{(2)}=\{(0,1),(1,1)\}.$
%    \begin{align}
%        \mc{D}^{(1)}&=\{(1,0),(1,1)\},\\
%        \mc{D}^{(2)}&=\{(0,1),(1,1)\}.
%    \end{align}
    When $T=3$, we have %$\mc{D}^{(1)}=\{(1,0,0), (1,0,1), (1,0,2),(1,1,0), (1,1,1)\}$, $\mc{D}^{(2)}=\{(0,1,0), (0,1,1), (1,1,0), (1,1,1)\}$, and $\mc{D}^{(3)}=\{(0,0,1), (0,1,1), (1,0,1), (1,1,1)\}.$
    \begin{align}
        \mc{D}^{(1)}&=\{(1,0,0), (1,0,1), (1,0,2),(1,1,0), (1,1,1)\}, \\
        \mc{D}^{(2)}&=\{(0,1,0), (0,1,1), (1,1,0), (1,1,1)\}, \\
        \mc{D}^{(3)}&=\{(0,0,1), (0,1,1), (1,0,1), (1,1,1)\}.
    \end{align}
\end{example}
The decoding process performed at the end of a frame then proceeds as follows. For every slot $n\in [N]$ during a frame, the corresponding transmission pattern $\mathbf{c}_n$ is computed\footnote{The transmission patterns $\mathbf{c}_n$ can evolve with iteration; however, we ignore the index for iterations to avoid heavy notation.}. The intra-slot SIC is then employed to resolve a type $t$ packet if there is a $\mathbf{c}_n$ for $n\in[N]$ belonging to the type $t$ decodable set $\mc{D}\upt$. We assume that in the burst payload of each packet, there is information about the (other) slots containing copies of this packet. Hence, the decoded packets are then used to cancel other replicas via inter-slot SIC. This completes an iteration and the decoding process continues iteratively until no more packets are decodable for all $t\in[T]$ and all $n\in[N]$ or all the packets in this frame are decoded. This process is done, independently, for every frame. }

\begin{figure}
    \centering
    \includegraphics[width=2.3in]{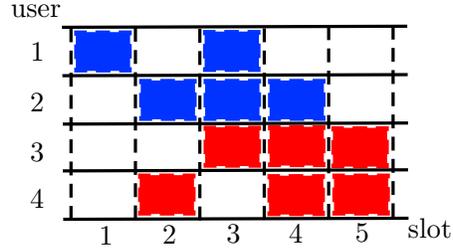}
    \caption{An example of SA with heterogeneous users transmitting in a frame of 5 slots. Here, we have $T=2$ types where the first type consists of users 1 and 2 and the second type consists of users 3 and 4.}
    \label{fig:ALOHA_ex}
\end{figure}

An example of the proposed IRSA with NOMA for $T=2$ can be found in Fig.~\ref{fig:ALOHA_ex}. In this figure, the replicas of packets transmitted from the type 1 users, namely users 1 and 2, are colored in blue and that from the type 2 users, namely users 3 and 4, are colored in red. In this example, user 1's packet can be easily decoded from slot 1. Subtracting user 1's packet at slot 3 from the decoded packet, there's no more degree-1 slot and traditional IRSA would stop the decoding process. For the proposed IRSA with NOMA, at slot 2, we receive the collision of a packet from type 1 (user 2) and another one from type 2 (user 4), both packets can be decoded via intra-slot SIC. Hence, our proposed IRSA with NOMA would continue the decoding process by subtracting users 2 and 4's packets in slot 4 and decode user 4's packet. %We would like to emphasize that unlike multi-packet reception, the proposed protocol cannot directly decode the packets in slot 5 via intra-slot SIC as they belong to the same type.

At the end of the decoding process, suppose the base station can successfully decode $\hat{K}^{(t)}$ packets of type $t\in[T]$, we define the (\textit{actual}) efficiency of type $t\in[T]$ achieved by the scheme as
\begin{equation}\label{eqn:true_eff}
    \hat{\eta}^{(t)} = \frac{\hat{K}^{(t)}}{N},\quad t\in[T],
\end{equation}
and the (\textit{actual}) sum efficiency of the scheme as
%\begin{equation}
   $ \hat{\eta} = \sum_{i=1}^T \hat{\eta}^{(t)}.$
%\end{equation}
It is clear that choosing different $\mc{L}^{(t)}$ would result in different efficiency. Throughout the paper, we call a collection of $T$ degree distributions $\{\mc{L}^{(1)}, \mc{L}^{(2)}, \ldots, \mc{L}^{(T)}\}$, one for each type $t\in[T]$, a policy. The goal of this paper is then to design the policy, such that the achieved efficiency is maximized.

%\begin{remark}
%    It is worth noting that the heterogeneity mentioned in this section and considered throughout the paper is the heterogeneity among different users/devices within the same use scenario of mMTC. It should not be confused with the heterogeneity among devices in different use scenarios such as that discussed in \cite{popovski_slicing18}.
%\end{remark}

\subsection{Comparison with existing models}
In this subsection, we discuss differences between the model considered in this paper and existing models.
\begin{remark}
    It is not difficult to see that when $T=1$, i.e., there is only one type of homogeneous users, the proposed protocol reduces to the IRSA protocol in \cite{Liva2011:IRSA_TCOM}. In light of this, the proposed protocol can be regarded as a generalization of the IRSA protocol to accommodate heterogeneous types of users.
\end{remark}

\begin{figure}
    \centering
    \includegraphics[width=3.2in]{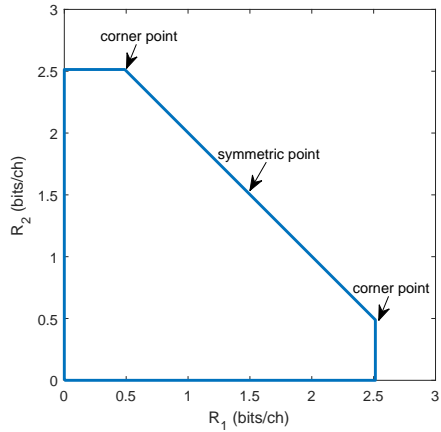}
    \caption{The capacity region of a two-user multiple access channel.} %In multi-packet reception, the base station has to be able to achieve the symmetric rate point with the same code as the users are homogeneous. In power-domain NOMA, exploiting heterogeneity among users, the base station uses intra-slot SIC to achieve a corner point.}
    \label{fig:Capacity region}
\end{figure}

\begin{remark}\label{rmk:NOMAvsMPR}
    We note that both multi-packet reception \cite{verdu_MPR88} and power-domain NOMA \cite{mazzini98, Dai15, Saito13VtcS, Saito13Pimrc} are based on the same physical-layer model, namely the multiple access channel \cite{cover91}. However, the multi-packet reception technique treats all the users equally and assumes collided packets are decodable as long as the number of users collided at a slot is smaller than a predefined number. Hence, in this homogeneous setting, the problem in the physical layer becomes to establish reliable multiple access with homogeneous users that have the same codebook and code rate. To the best of our knowledge, this is not an easy task that is tackled with some success in \cite{krishna11_MACviaSC} and \cite{Zhu17}, which involves either extremely large blocklength due to spatial coupling \cite{krishna11_MACviaSC} or very complex operations of coding over prime fields\footnote{This is due to the adoption of Construction A lattices \cite{erez04}.}. On the contrary, the power-domain NOMA lets the users transmit with different power levels and/or rates so that low-complex SIC can be employed to achieve a corner point of the capacity region\footnote{\textcolor{black}{Here, the standard information-theoretic sense of achievability \cite{cover91} is considered.}}. An illustration can be found in Fig.~\ref{fig:Capacity region}. In this paper, the considered problem exhibits natural heterogeneity among users; hence, it is more natural and suitable to work with power-domain NOMA so that the heterogeneity can be exploited by intra-SIC. The main theme of this paper then centers around analyzing policies for IRSA with power-domain NOMA and designing policies that optimally exploit the heterogeneity inherent in the considered problem.
\end{remark}

\begin{remark}
    The combination of NOMA and SA has been investigated before in \cite{choi17, choi18} by Choi. However, in these works, it is assumed that there is only one type of homogeneous users. The focus of \cite{choi17} is mainly to analyze the performance of SA when users randomly choose their power levels and the receiver employs intra-slot SIC. Moreover, in both works \cite{choi17, choi18}, the inter-slot SIC technique is not considered.
    %In \cite{choi18}, a game-theoretic approach is taken where each user is regarded as a player and a payoff function closely related to energy efficiency is adopted. Under this framework, the author derives a mixed strategy Nash equilibrium where no user would have incentive to unilaterally deviate from this strategy. Moreover, in both works \cite{choi17, choi18}, the inter-slot SIC technique is not considered. At this point, it is quite clear that although both the present work and \cite{choi17, choi18} consider NOMA, the motivations, goals, and approaches are radically different.
\end{remark}

\begin{remark}
    We also note that our model is fundamentally different from IRSA with capture effect discussed in \cite[Appendix A]{Liva2011:IRSA_TCOM} and \cite{Stefanovic17}. With capture effect, it is assumed that the receiver may be able to decode more than one packet with some probability. This opportunity usually comes from the effect of random channel fading. On the other hand, in the considered model, the receiver is able to decode more than one packet {\it deterministically} for certain configurations of collided users, due to the natural heterogeneity inherent in the problem.
\end{remark}

\subsection{Implementation Issues}
We now address potential implementation issues of the proposed IRSA with NOMA. We would like to note that many of these issues have been encountered and addressed by the conventional IRSA schemes \cite{Liva2011:IRSA_TCOM}. However, we still present them here for the sake of self-containedness.
\begin{enumerate}
  \item {\it Slot synchronization:} This issue is inherent in all slotted multiple access schemes. One easy solution to provide slot synchronization is to rely on stable clocks and a small amount of guard time between packets \cite[Sec. 4.2]{BertsekasGallager92}. \textcolor{black}{Moreover, for applications with cheap user devices, stable clocks may be unaffordable. For such scenarios, \cite{Polonelli18} provides a technique to achieve slot synchronization, where each node constantly re-synchronizes its clock with the base station by exploiting the ACK signals fed back from the base station. Another source of slot synchronization errors which is less seen in traditional communication systems is from significantly different travel distances experienced by different devices' signals. This type of source of slot synchronization errors has become important in non-terrestrial networks (NTN) \cite{TR38.821NTN}. Fortunately, in NTN, each device is required to be GNSS\footnote{GNSS stands for Global Navigation Satellite System.}-capable for acquiring its own position and together with satellite ephemeris. Each user can calculate and then pre-compensate its timing advance (TA) (see Section 6.3 in \cite{TR38.821NTN}). To reduce devices' computation loading, another candidate technique currently under discussion in 3GPP is that the base station estimates and broadcasts TA to a group of devices who have similar TA (see Section 6.3 in \cite{TR38.821NTN}).}

  \item {\it Frame synchronization:} This issue is also not new and has to be cope with in every frame-synchronous systems. This issue can be addressed by letting the base station periodically broadcasting a beacon signal at the beginning of each frame. In Section~\ref{sec:FA}, extension to frame asynchronous setting will be discussed.
  \item {\it Inter-slot SIC:} Similar to CRDSA \cite{Casini:CRDSA} and IRSA schemes \cite{Liva2011:IRSA_TCOM}, to enable inter-slot SIC, the proposed protocol requires the knowledge of the locations of other replicas once a packet is successfully decoded. \textcolor{black}{This can be achieved by including pointers to the locations of other replicas into the burst payload or by some pseudo-random mechanism known by both the transmit and receive ends. For example, the DVB system \cite{DVB14} adopts the latter one, where each active user reports its MAC address and a pseudo-random number to the base station. Both the base station and the user can then compute the slots used for sending replicas by a common deterministic function of the MAC address and the pseudo-random number.}
  \item {\it Intra-slot SIC:} \textcolor{black}{For the proposed scheme, in order to enable intra-slot SIC at a slot, it appears that the receiver has to know which types of packets are collided in this slot.} This issue is new in our proposed protocol. \textcolor{black}{We provide two ways to address this issue. We note that in DVB-RCS2 \cite{DVB14}, each active user selects the number of transmissions and a random seed for uniquely determining a pseudo random pattern. This information is then stored in the burst payload and sent to the base station. Given the existing data structure, to enable intra-slot SIC, each user can include additional $\log(T)$ bits to the payload for revealing its type. The second method described in the sequel eliminates the need for sending extra $\log(T)$ bits by increasing the decoding burden.}
      Let $R^{(t)}$ be the rate of the physical-layer channel code adopted by users of type $t\in[T]$. Let us assume the We assume that the rate tuple $(R^{(1)}, R^{(2)}, \ldots, R^{(T)})$ can be achieved by intra-slot SIC with ascending decoding order $1\rightarrow 2\rightarrow\ldots\rightarrow T$, without loss of generality. Following the NOMA principle, a means to address this issue is as follows: Starting from $t=1$, the base station tries to decode a packet of type $t$; if it succeeds\footnote{The success of decoding can be checked by a cyclic redundancy check mechanism.}, the decoded packet is subtracted from the received signal and we set $t=t+1$; otherwise, directly set $t=t+1$ and look for decoding opportunity of the next type. This procedure would continue until $t=T$.
  %\item {\it Estimation of the number of active users:} To obtain near optimal performance, the knowledge of the number of active users is required in most of the SA-based UMA schemes. This issue can be cope with by the technique in \cite{Stefanovic13_user_estimate}.
\end{enumerate}

\section{Graph Representation}\label{sec:graph}
In this section, we introduce a graph representation of the proposed IRSA with NOMA protocol for determining an optimal policy in Section~\ref{sec:DE}.

\subsection{Bipartite Graph and Degree Distributions}
\begin{figure}
    \centering
    \includegraphics[width=1.8in]{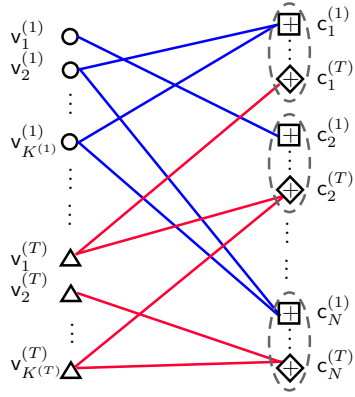}
    \caption{A bipartite graph representation of the proposed IRSA with NOMA.}
    \label{fig:IRSA_NOMA_graph}
\end{figure}

We first note that in UMA with $T$ types of heterogeneous users, each type has its own transmission policy and number of users that may be different from other types. We construct a bipartite graph representation of the proposed IRSA with NOMA protocol. The graph consists of $\sum_{t=1}^T K\upt$  \textit{variable nodes} and $N$ \textit{super check nodes}, as shown in Fig.~\ref{fig:IRSA_NOMA_graph}. Each variable node $\Vt_j$ represents a user $j\in[K^{(t)}]$ of type $t\in[T]$. Each super check node consists of $T$ check nodes $\Ct_n$, one for each type $t\in[T]$. A user $j$ of type $t\in[T]$ transmits a packet in slot $n\in[N]$ if and only if there is an edge connecting $\Vt_j$ with $\Ct_n$. For simplicity, we use $\Vt$ and $\Ct$ to denote a generic variable and check nodes of type $t\in[T]$.
%Thus, in our graph representation shown in Fig.~\ref{fig:IRSA_NOMA_graph}, there are $T$ types of variable nodes, where $\Vt_j$ stands for user $j\in[K^{(t)}]$ of type $t\in[T]$. For the ease of presentation, we denote by $\Vt$ a generic user of type $t\in[T]$. Also, we use $N$ super check nodes to denote the $N$ time slots. The super check node $n\in[N]$ consists of $T$ types of check nodes $\Ct_n$, $t\in[T]$, one for each type of users. Again, we use $\Ct$ to denote a generic check node of type $t\in[T]$. A user $j$ of type $t\in[T]$ transmits a packet in slot $n\in[N]$ if and only if there is an edge connecting $\Vt_j$ with $\Ct_n$.

With this bipartite graph representation, a user of type $t$ who transmits $d$ replicas corresponds to a degree $d$ variable node $\Vt$. Moreover, a time slot in which $d$ type $t$ variable nodes transmit corresponds to a degree $d$ check node $\Ct$. We define the node perspective left and right degree distributions as
\begin{equation}\label{eqn:node_dist}
    \mc{L}\upt[x] = \sum_{d} L\upt_d x^d \quad\text{and}\quad \mc{R}\upt[x] = \sum_d R\upt_d x^d,
\end{equation}
respectively, where $L\upt_d$ is the probability that a variable node $\Vt$ has degree $d$ and $R\upt_d$ is the probability that a check node $\Ct$ has degree $d$. Through the node perspective degree distributions, the edge perspective left and right degree distributions can be given by
\begin{align}\label{eqn:edge_dist_left}
    \lambda\upt[x] = \sum_d \lambda\upt_d x^{d-1}=\frac{\mc{L}^{'(t)}[x]}{\mc{L}^{'(t)}[1]},
\end{align}
and
\begin{align}\label{eqn:edge_dist_right}
   \rho\upt[x] &= \sum_d\rho\upt_d x^{d-1} =\frac{\mc{R}^{'(t)}[x]}{\mc{R}^{'(t)}[1]},
\end{align}
respectively, where $\lambda\upt_d$ stands for the probability that an edge connects to a type $t$ variable node of degree $d$ and $\rho\upt_d$ is the probability that an edge connects to a type $t$ check node of degree $d$.

The \textit{target} efficiency of type $t\in[T]$ is then defined as
\begin{equation}
    \eta\upt = \frac{K\upt}{N} = \frac{\mc{R}^{'(t)}[1]}{\mc{L}^{'(t)}[1]},
\end{equation}
and the \textit{target} sum efficiency of the scheme is defined as
%\begin{equation}
$    \eta = \sum_{t=1}^T \eta\upt.$
%\end{equation}
Note that the target efficiency $\eta\upt$ is different from the actual efficiency $\hat{\eta}\upt$ in \eqref{eqn:true_eff}. In general, $\hat{\eta}\upt\leq \eta\upt$ with equality if and only if all packets are successfully decoded.

\begin{figure}
    \centering
    \includegraphics[width=3.2in]{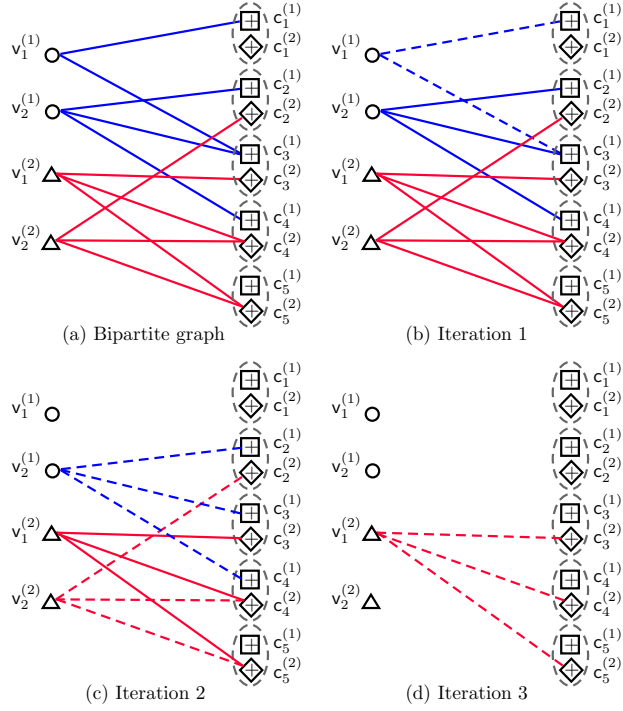}
    \caption{The graph representation of the scheme in Fig.~\ref{fig:ALOHA_ex} and its decoding output at each iteration of the modified peeling decoder.}
    \label{fig:IRSA_NOMA_ex}
\end{figure}

\begin{example}\label{ex:graph_peel}
    Consider the scheme described in Fig.~\ref{fig:ALOHA_ex} where there is one type 1 variable node of degree 2 and one type 1 variable node of degree 3. Hence, $\mc{L}^{(1)}[x]=0.5x^2 +0.5x^3$. One can obtain $\mc{L}^{(2)}[x]=x^3$ in a similar fashion. Also, there are three type 1 check nodes of degree 1 and one type 1 check node of degree 2. Therefore, $\mc{R}^{(1)}[x]=0.75x+0.25x^2$. Similarly, $\mc{R}^{(2)}[x]=0.5x+0.5x^2.$ Moreover, one can easily verify that the edge perspective degree distributions are given by
    \begin{align}
        \lambda^{(1)}[x]= \frac{2}{5}x + \frac{3}{5}x^2, \quad \lambda^{(2)}[x]=x^2, \quad \rho^{(1)}[x]=\frac{3}{5}+\frac{2}{5}x,\quad \rho^{(2)}[x]=\frac{1}{3}+\frac{2}{3}x.
    \end{align}
    The graph representation of this example can be found in Fig.~\ref{fig:IRSA_NOMA_ex}-(a).
\end{example}

\subsection{Modified Peeling Decoder}
Here, we propose a modified peeling decoder for the bipartite graph that corresponds to the decoding procedure of the proposed IRSA with NOMA. The decoder operates in an iterative fashion. For each iteration, the decoder looks at every super check node $n\in[N]$ and computes the remaining degrees of check nodes $\Ct_1, \Ct_2, \ldots, \Ct_T$, which are stored in the transmission pattern $\mathbf{c}_n$. We then look for an $n$ such that $\mathbf{c}_n$ belongs to the type $t$ decodable set $\mc{D}\upt$ for some $t\in[T]$. If such an $n$ exists, we use intra-slot SIC to decode all the type $t$ packets satisfying $\mathbf{c}_n\in\mc{D}\upt$ in this super check node $n$. For a variable node $\Vt$ that is decoded, all the edges connected with it are removed from the graph. This concludes an iteration. The modified peeling decoder operates iteratively until no more decoding opportunity is found.
\addtocounter{lemma}{-1}
\begin{example}[continued]
    In this example, we illustrate how to use the modified peeling decoder to decode the scheme in Example~\ref{ex:graph_peel}. As shown in Fig.~\ref{fig:IRSA_NOMA_ex}-(b), in iteration 1, $\msf{c}^{(1)}_1$ has degree 1 and $\msf{c}^{(2)}_1$ has degree 0, i.e., $\mathbf{c}_1=(1,0)\in\mc{D}^{(1)}$; hence, $\msf{v}^{(1)}_1$ is decoded and all its edges are removed. Then, in iteration 2 shown in Fig.~\ref{fig:IRSA_NOMA_ex}-(c), $\msf{c}^{(1)}_2$ has degree 1 and $\msf{c}^{(2)}_2$ has degree 1, i.e., $\mathbf{c}_2=(1,1)$ belonging to both $\mc{D}^{(1)}$ and $\mc{D}^{(2)}$; thus, both $\msf{v}^{(1)}_2$ and $\msf{v}^{(2)}_2$ can be decoded by intra-slot SIC. All the edges connected to $\msf{v}^{(1)}_2$ and $\msf{v}^{(2)}_2$ are subsequently removed. Last, in iteration 3 shown in Fig.~\ref{fig:IRSA_NOMA_ex}-(d), one can easily see that $\msf{v}^{(2)}_1$ can be decoded. %Here, we would like to reiterate that unlike multi-packet reception, the collision in the super check node  $5$ cannot be resolved at the iteration 1 since $\msf{c}^{(2)}_5$ has degree 2.
\end{example}

\section{Multi-Dimensional Density Evolution and Convergence Analysis}\label{sec:DE}
In this section, to analyze the IRSA with NOMA, we propose a novel multi-dimensional density evolution based on the graph representation described in Section~\ref{sec:graph}. With this multi-dimensional density evolution, we then provide the convergence analysis of the IRSA with NOMA under modified peeling decoding. With an assist from the convergence analysis, we formulate and numerically solve an optimization problem that finds best left degree distributions, leading to the highest asymptotic efficiency. For clearly explaining the ideas behind our scheme and analysis, we focus on $T=2$. The general $T$ case can be similarly analyzed. %See Appendix~\ref{apx:T_type_DE} for the general $T$. %\textcolor{red}{Throughout the section, we focus on $T=2$ for the sake of conciseness and leave the general case to Appendix~\ref{apx:T_type_DE}.}

\subsection{Proposed Multi-Dimensional Density Evolution}\label{subsec:density_evo}
It is well known that for LDPC codes over a binary erasure channel, BP and peeling decoders have the same performance\footnote{It essentially means that the order of the limit of average residue erasure probability as the number of iterations tends to $\infty$ and the limit of that as the blocklength tends to $\infty$ is exchangeable. The interested reader is referred to \cite[Sec. 3.19]{urbanke_book}}\footnote{The reason that we propose using the peeling decoder instead of BP is because the peeling decoder potentially leads to a smaller decoding latency in practice as it does not have to wait until the end of the frame in order to start decoding; it can start decoding right away once a decodable slot shows up.}. Moreover, density evolution is an outstanding tool for analyzing the performance of BP decoding over a sparse graph \cite{urbanke_book}. Thus, this section proposes a novel multi-dimensional density evolution to analyze the performance of the proposed scheme under BP decoding. \textcolor{black}{Similar to that introduced in \cite[Sec. 2.5]{urbanke_book}, the BP decoding is an iterative algorithm where in each iteration, each variable node computes the belief of its own message and passes an extrinsic version of it to each connected check node, while each check node computes its belief of the message of each connected variable node and passes an extrinsic version to that variable node. After a predefined number of iterations, the decoding algorithm halts and outputs hard decisions according to the last beliefs held by the variable nodes.} %Simulation results in Section~\ref{sec:simulation} will confirm the accuracy of the proposed analysis for the proposed scheme with the modified peeling decoder.

\textcolor{black}{Consider BP decoding described above.} For $t\in[2]$, we denote by $x_\ell\upt$ and $y_\ell\upt$ the average erasure probability of the message passed along an edge from $\Vt$ to $\Ct$ and that of the message passed along an edge from $\Ct$ to $\Vt$ in iteration $\ell$, respectively. At iteration $\ell=0$, according to the decodable set $\mc{D}\upt$ specified in Example~\ref{ex:decodable_set}, for an edge connected to a type $t$ check node $\Ct$, the outgoing message is not in erasure if and only if a) $\Ct$ has degree 1; b)  $\msf{c}^{(\bar{t})}$ has a degree either 0 or 1 for every $\bar{t}\in[2],~\bar{t}\neq t$. Hence, we initialize $y_0\upt$ for initial iteration $\ell=0$ to be
\begin{equation}
    y_0\upt = 1-\rho\upt[0]\cdot \left( \mc{R}^{(\bar{t})}[0] + \mc{R}^{'(\bar{t})}[0] \right),
\end{equation}
where $\mc{R}^{(\bar{t})}[0]=R^{(\bar{t})}_0$ and $\mc{R}^{'(\bar{t})}[0]=R^{(\bar{t})}_1$ are the fractions of type $\bar{t}$ check nodes having degree 0 and degree 1, respectively. Moreover, the term $\rho\upt[0]=\rho_1\upt$ is the probability that an edge connects to a type $t$ check node of degree $1$.

Suppose we have obtained $x_\ell\upt$ and $y_\ell\upt$ for all $t\in[T]$ from iteration $\ell$. In iteration $\ell+1$, for an edge incident to a variable node $\Vt$ with degree $d$, the only possibility that the message along this edge to a $\Ct$ is in erasure is that all the other $d-1$ edges are in erasure. Therefore, the probability that the message passed along this edge is in erasure is $(y_\ell\upt)^{d-1}$. Now, averaging over all the edges results in the average erasure probability
\begin{equation}\label{eqn:DE_x}
    x_\ell\upt = \sum_d \lambda\upt_d (y_\ell\upt)^{d-1} = \lambda\upt [y_\ell\upt ].
\end{equation}

For an edge incident to a check node $\Ct$ with degree $d$, the message passed along this edge to a $\Vt$ is not in erasure if and only if a) all the other $d-1$ edges incident to $\Ct$ are not in erasure in the previous iteration; and b) all but at most one of the edges incident to $\msf{c}^{(\bar{t})}$ are not in erasure in the previous iteration for every $\bar{t}\in[2],~\bar{t}\neq t$. Suppose in the same super check node $n$, $\msf{c}^{(\bar{t})}_n$ has degree $d_{\bar{t}}$. Then the above event has probability
\begin{equation}
    (1-x_\ell\upt)^{d-1}\cdot\left( (1-x_\ell^{(\bar{t})})^{d_{\bar{t}}} + d_{\bar{t}} x_\ell^{(\bar{t})}(1-x_\ell^{(\bar{t})})^{d_{\bar{t}}-1}\right),
\end{equation}
where $(1-x_\ell^{(\bar{t})})^{d_{\bar{t}}}$ is the probability that all $d_{\bar{t}}$ edges are not erased and $d_{\bar{t}} x_\ell^{(\bar{t})}(1-x_\ell^{(\bar{t})})^{d_{\bar{t}}-1}$ is the probability that all but one edges are not erased. Now, averaging over all the edges and over all the type $\bar{t}$ check nodes for $\bar{t}\in[2],~\bar{t}\neq t$ shows that the average probability of correct decoding is given by
\begin{align}\label{eqn:DE_y_intermediate}
    &\sum_d \lambda\upt_d(1-x_\ell\upt)^{d-1}\cdot\nonumber  \sum_{d_{\bar{t}}}R^{(\bar{t})}_{d_{\bar{t}}}\left( (1-x_\ell^{(\bar{t})})^{d_{\bar{t}}} + d_{\bar{t}} x_\ell^{(\bar{t})}(1-x_\ell^{(\bar{t})})^{d_{\bar{t}}-1}\right) \nonumber \\
    &=\rho\upt[1-x_\ell\upt] \left( \mc{R}^{(\bar{t})}[1-x_\ell^{(\bar{t})}] + x_\ell^{(\bar{t})}\mc{R}^{'(\bar{t})}[1-x_\ell^{(\bar{t})}]\right).
\end{align}
Therefore, the average erasure probability becomes
\begin{align}\label{eqn:DE_y}
    y_{\ell+1}\upt = 1-\rho\upt[1-x_\ell\upt]\cdot  \left( \mc{R}^{(\bar{t})}[1-x_\ell^{(\bar{t})}] + x_\ell^{(\bar{t})}\mc{R}^{'(\bar{t})}[1-x_\ell^{(\bar{t})}]\right).
\end{align}
Plugging \eqref{eqn:DE_x} into \eqref{eqn:DE_y} leads to the evolution of average erasure probability of a type $t$ check node as shown in \eqref{eqn:DE_full} in the bottom of next page.
\begin{figure*}[!b]
\normalsize
\hrulefill
\begin{IEEEeqnarray}{rCl}\label{eqn:DE_full}
\hspace{-25pt}     y_{\ell+1}\upt=1-\rho\upt[1-\lambda\upt[y_\ell\upt]]\cdot \left( \mc{R}^{(\bar{t})}[1-\lambda^{(\bar{t})} [y_\ell^{(\bar{t})}] ] + \lambda^{(\bar{t})}[y_\ell^{(\bar{t})}]\mc{R}^{'(\bar{t})}[1-\lambda^{(\bar{t})}[y_\ell^{(\bar{t})}]]\right),\quad t\in[2].
\end{IEEEeqnarray}
%\setcounter{equation}{\value{tempeqcounter}} % restore correct value
%\vspace*{4pt}
\end{figure*}

\subsection{Convergence and Stability Condition}
After obtaining the density evolution in \eqref{eqn:DE_full}, for any given degree distributions $\mc{L}\upt$ (or $\lambda\upt$) and $\mc{R}\upt$ (or $\rho\upt$), one can now analyze whether {\color{black} the average erasure probability converges to 0 by checking whether $y_\ell\upt> y_{\ell+1}\upt$ for every $\ell$, starting from $y_0\upt=1$ for $t\in[T]$.}
%if $(0,0)$ is the only fixed-point solution within $[0,1]\times [0,1]$ to \eqref{eqn:DE_full}. It suffices to check $y_\ell\upt> y_{\ell+1}\upt$ for every $\ell$ and every $y_\ell\upt>0$.

Moreover, to make sure that the average erasure probability indeed vanishes instead of hovering around 0, we derive the following stability condition\footnote{\textcolor{black}{Note that the term ``stability" used here is referred to the stability of fixed points rather than that in the stability analysis of slotted ALOHA \cite{Tsybaov79, BertsekasGallager92}.}}. We enforce $\lambda_1\upt=0$ for all $t\in[2]$ because we certainly do not want degree 1 variable nodes. Then, assuming $y\upt$ is very small for all $t\in[2]$, we expand the degree distributions and approximate them by keeping only the linear terms as follows,
\begin{align}
    \lambda\upt[y\upt] &\approx \lambda_2\upt y\upt, \label{eqn:stable_1} \\
    \rho\upt[1-\lambda\upt[y\upt]] &\approx 1-\rho^{'{(t)}}[1]\lambda_2\upt y\upt, \label{eqn:stable_2} \\
    R^{(\bar{t})}[1-\lambda^{(\bar{t})}[y^{(\bar{t})}]] &\approx 1-R^{'(\bar{t})}[1] \lambda_2^{(\bar{t})} y^{(\bar{t})}, \label{eqn:stable_3} \\
    R^{'(\bar{t})}[1-\lambda^{(\bar{t})}[y^{(\bar{t})}]] &\approx R^{'(\bar{t})}[1] - R^{''(\bar{t})}[1]\lambda_2^{(\bar{t})}y^{(\bar{t})}.\label{eqn:stable_4}
\end{align}
We can now linearize the recursion around 0 by plugging \eqref{eqn:stable_1}-\eqref{eqn:stable_4} into \eqref{eqn:DE_full} to get
\begin{align}
    y\upt > 1- (1-\rho^{'{(t)}}[1]\lambda_2\upt y\upt) \left(1-R^{''(\bar{t})}[1](\lambda_2\upt y^{(\bar{t})})^2\right) \approx \rho^{'{(t)}}[1]\lambda_2\upt y\upt,
\end{align}
which leads to the following stability condition
\begin{equation}\label{eqn:stable_con}
    \lambda_2\upt < \frac{1}{\rho^{'{(t)}}[1]}\quad\text{for $t\in[2]$}.
\end{equation}

\subsection{Optimization Problem}\label{subsec:opt_problem}
Before we formulate the optimization problem, we note that there are two sets of degree distributions $\{\mc{L}\upt[x]\}$ and $\{\mc{R}\upt[x]\}$ in the proposed problem, but we only have control over $\{\mc{L}\upt[x]\}$. The behavior of $\{\mc{R}\upt[x]\}$; however, are completely determined by how the users behave. Specifically, according to the protocol, a type $t$ user $\Vt$ having degree $L$ will send a replica in slot $\Ct$ with probability $L/N$. Hence the average probability that a user $\Vt$ sending a replica in slot $\Ct$ is given by
%\begin{equation}
$    \frac{\mc{L}^{'(t)}[1]}{N} = \frac{\mc{R}^{'(t)}[1]}{K\upt}.$
%\end{equation}
Thus, the degrees of $\Ct$ follows the Binomial distribution with parameter $\frac{\mc{L}^{'(t)}[1]}{N}$. Similar to \cite{Liva2011:IRSA_TCOM}, by Poisson approximation \cite{Mitzenmacher_book05}, we have
\begin{align}\label{eqn:poisson_right}
    \mc{R}\upt[x] \approx \exp\left( -\mc{R}^{'(t)}[1](1-x) \right) = \exp\left( -\eta\upt \mc{L}^{'(t)}[1](1-x) \right).
\end{align}
Moreover, from \eqref{eqn:edge_dist_left}, we have
\begin{equation}\label{eqn:poisson_approx_proposed}
    \rho\upt[x]= \exp\left( -\eta\upt \mc{L}^{'(t)}[1](1-x) \right).
\end{equation}
\textcolor{black}{Now, the convergence condition becomes \eqref{eqn:DE_full_poisson} in the bottom of the this page by plugging \eqref{eqn:poisson_approx_proposed} into \eqref{eqn:DE_full} and noting that it is sufficient that $y^{(t)}_{\ell+1}<y^{(t)}_{\ell}$ for every $\ell$ with $y^{(t)}_{\ell}\neq 0$.} %With this, the convergence condition in \eqref{eqn:DE_full} becomes \eqref{eqn:DE_full_poisson} in the bottom of the this page.
\begin{figure*}[!b]
\normalsize
\hrulefill
\begin{align}\label{eqn:DE_full_poisson}
     &\hspace{-10pt}y_\ell\upt>1-\exp\left( -\eta\upt \mc{L}^{'(t)}[1] \lambda\upt[y_\ell\upt]\right) \cdot \nonumber \\
     &\hspace{-10pt} \left( \exp\left( -\eta^{(\bar{t})} \mc{L}^{'(\bar{t})}[1] \lambda^{(\bar{t})} [y_\ell^{(\bar{t})}] \right)+  \eta^{(\bar{t})} \mc{L}^{'(\bar{t})}[1] \lambda^{(\bar{t})} [y_\ell^{(\bar{t})}] \exp\left( -\eta^{(\bar{t})} \mc{L}^{'(\bar{t})}[1] \lambda^{(\bar{t})} [y_\ell^{(\bar{t})}] \right)\right),\quad t\in[2].
\end{align}
%\setcounter{equation}{\value{tempeqcounter}} % restore correct value
%\vspace*{4pt}
\end{figure*}

%\begin{figure*}[!b]
%\normalsize
%%\setcounter{tempeqcounter}{\value{equation}} % temp store of current value
%\hrulefill
%\begin{IEEEeqnarray}{rCl}\label{eqn:DE_full_poisson}
%%\setcounter{equation}{\value{storeeqcounter}} % number of this equation
%     y_{\ell+1}\upt&=&1-\exp\left( -\eta\upt \mc{L}^{'(t)}[1] \lambda\upt[y_\ell\upt]\right) \cdot \nonumber \\
%     && \prod_{\bar{t}=1, \bar{t}\neq t}^T \left( \exp\left( -\eta^{(\bar{t})} \mc{L}^{'(\bar{t})}[1] \lambda^{(\bar{t})} [y_\ell^{(\bar{t})}] \right)+  \eta^{(\bar{t})} \mc{L}^{'(\bar{t})}[1] \lambda^{(\bar{t})} [y_\ell^{(\bar{t})}] \exp\left( -\eta^{(\bar{t})} \mc{L}^{'(\bar{t})}[1] \lambda^{(\bar{t})} [y_\ell^{(\bar{t})}] \right)\right).
%\end{IEEEeqnarray}
%%\setcounter{equation}{\value{tempeqcounter}} % restore correct value
%%\vspace*{4pt}
%\end{figure*}

Now, we are ready to formulate the optimization problem that maximizes the target efficiency subject to conditions derived thus far:
\begin{subequations}
\begin{alignat*}{2}
&\!\max_{\{\mc{L}\upt[x]\} }        &\qquad& \eta =\sum_{t=1}^2 \eta\upt \\
&\text{subject to} &      & \text{convergence condition~}\eqref{eqn:DE_full_poisson},\\
%&                  &      & \text{stability condition~}\eqref{eqn:stable_con},\\
&                  &      & \lambda_1\upt=0~\text{for $t\in[2]$},\\
&                  &      & L_d\upt\geq 0, ~\text{for $d\in[\dmax\upt]$ and $t\in[2]$}, \\
&                  &      & \mc{L}\upt[1]=1~\text{for $t\in[2]$},
\end{alignat*}
\end{subequations}
where $\dmax\upt$ is the maximum degree of $\mc{L}\upt[x]$ that has to be imposed in practice. \textcolor{black}{We solve this problem and provide some optimized degree distributions in the next subsection. We note that the stability condition \eqref{eqn:stable_con} is not strictly required for maximizing the target efficiency as above. However, for applications that require very low packet loss rates, we do need to include \eqref{eqn:stable_con} into our optimization problem in order to make sure that the average erasure probability does vanish.}

%\begin{remark}\label{rmk:stability}
%    We would like to point out that if maximizing the target efficiency is the sole goal, then the stability condition \eqref{eqn:stable_con} is not strictly required and may be removed. However, for applications that require very low packet loss rates, we do need the stability condition in \eqref{eqn:stable_con} to make sure that the erasure probability does vanish.
%\end{remark}

\subsection{Solving the Optimization Problem}\label{subsec:solve_opt}
%Let us first review the optimization problem of the conventional IRSA (with only single type $T=1$). Let $y$ be the average erasure probability before an iteration and $y'$ be that after this iteration. The decodability condition of a left degree distribution for the conventional IRSA is simply to make sure that $y'<y$ for every iteration. A sufficient condition is then to check whether $y'<y$ for every $y\in(0,1]$. That is, whether the curve $y'$ against $y$, denoted as the evolution curve, lies entirely beneath the identity line $y'=y$. The optimization problem is then to find a degree distribution whose evolution curve almost, but not quite, touches the identity line. This problem is similar to design LDPC code ensemble and can then be solved by either linear programming \cite{Luby01peeling} or by differential evolution \cite{DE_book_price05}.

\begin{figure}
\begin{subfigure}{.5\textwidth}
    \centering
    \includegraphics[width=3.3in]{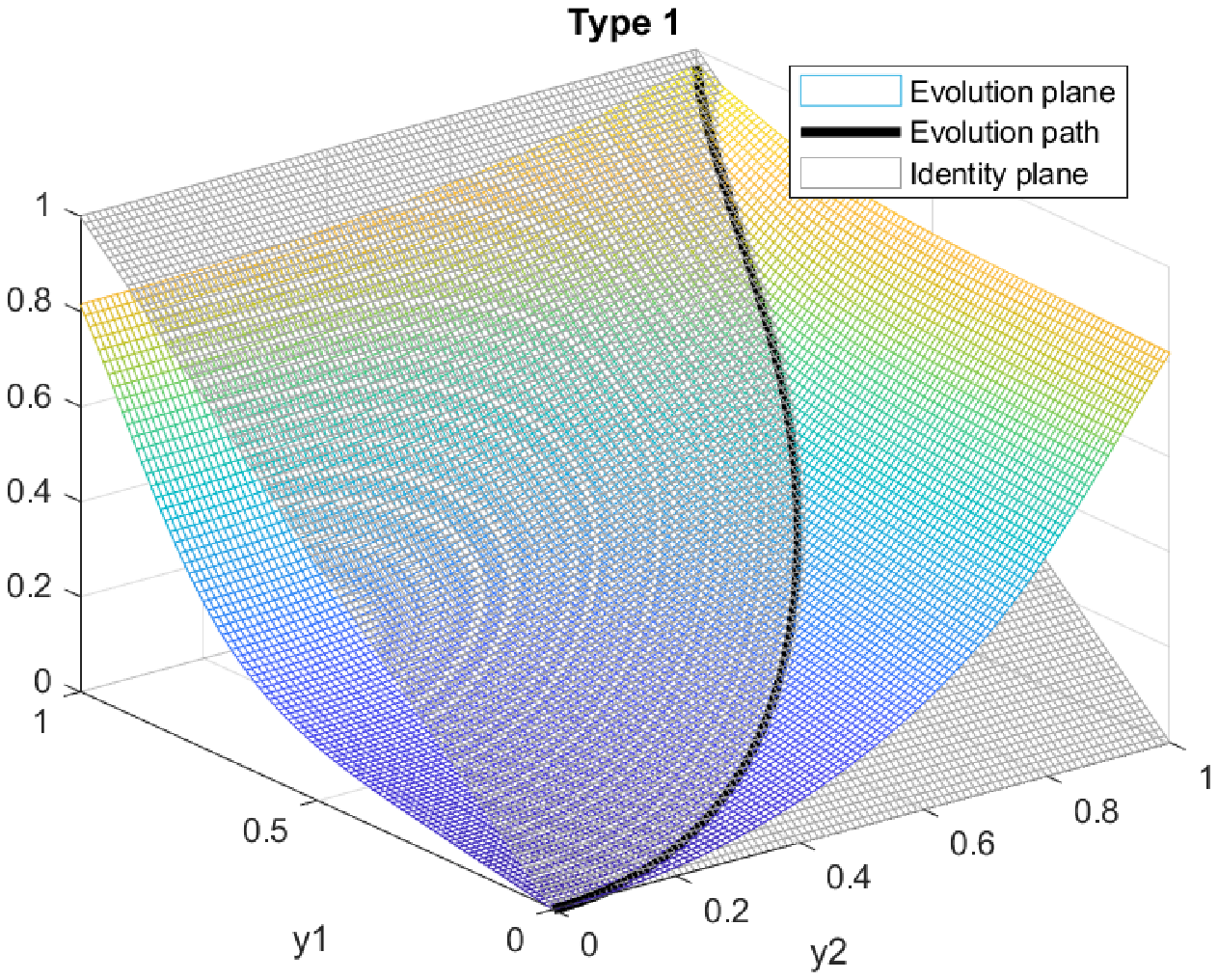}
    \caption{Type 1: $y^{(1)'}$.}
    \label{fig:evolution_plane_type1}
\end{subfigure}
\begin{subfigure}{.5\textwidth}
    \centering
    \includegraphics[width=3.3in]{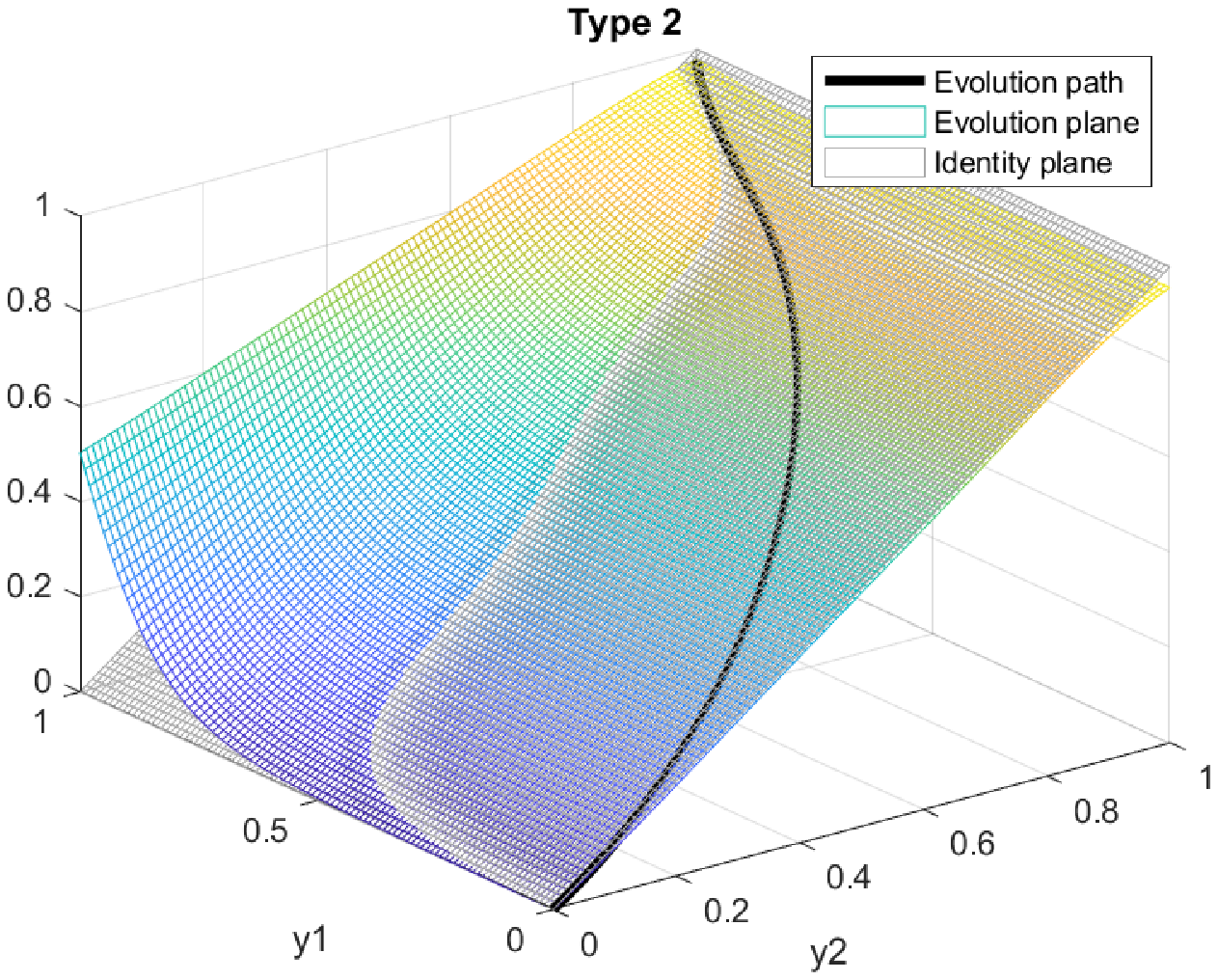}
    \caption{Type 2: $y^{(2)'}$.}
    \label{fig:evolution_plane_type2}
\end{subfigure}
\caption{Identity plane, evolution plane, and evolution path}
\label{fig:evolution}
\end{figure}

In what follows, we again focus on $T=2$ solely but the discussion and intuitions apply to any $T\geq 2$. For $T=2$, let $y\upt$ and $y^{(t)'}$ be \textcolor{black}{the LHS and RHS of \eqref{eqn:DE_full_poisson},} representing the average erasure probability before and that after an iteration, respectively. Similar to the conventional IRSA in \cite{Liva2011:IRSA_TCOM}, we can have a sufficient condition for decodability that $y^{(t)'}<y\upt$ for $t\in[2]$ and every $(y^{(1)},y^{(2)})\in(0,1]\times(0,1]$. This admits a graphical interpretation as follows. Note that it is a $T=2$ dimensional problem and we need a curve of $(y^{(1)'},y^{(2)'})$ against $(y^{(1)},y^{(2)})$, which is in 4 dimensional space. We instead plot $y^{(1)'}$ against $(y^{(1)},y^{(2)})$ and $y^{(2)'}$ against $(y^{(1)},y^{(2)})$ separately in Figs~\ref{fig:evolution_plane_type1} and \ref{fig:evolution_plane_type2}, respectively. \textcolor{black}{In Fig.~\ref{fig:evolution_plane_type1}, the identity plane is the hyperplane consisting of $(y^{(1)},y^{(2)},y^{(1)})$ and the evolution plane for the degree distribution pair $\mathsf{P}_2$ shown in Table~\ref{tbl:degree_dist} consists of $(y^{(1)},y^{(2)},y^{(1)'})$. Similarly, in Fig.~\ref{fig:evolution_plane_type2}, we show the identity plane consisting of $(y^{(1)},y^{(2)},y^{(2)})$ and also plot the evolution plane for $\mathsf{P}_2$ that consists of $(y^{(1)},y^{(2)},y^{(2)'})$.}
The sufficient condition mentioned above is then to ask the entire evolution plane lie beneath the identity plane in both figures. However, this is by no means necessary and is way too strong as after each iteration, both $y^{(1)}$ and $y^{(2)}$ drop and some pairs like $(0,1)$ and $(1,0)$ will never be visited. Also shown in Figs.~\ref{fig:evolution_plane_type1} and \ref{fig:evolution_plane_type2} is the evolution path of the considered distribution pair that is generated by using the output $(y^{(1)'},y^{(2)'})$ of the previous iteration as input. This evolution path depicts the trajectory of the pair of erasure probabilities evolving with the iterative decoding algorithm. Although the evolution plane does not lie entirely beneath the identity plane, since the evolution path in this example lies entirely beneath the identity plane and has the unique fixed point at the origin, this degree distribution pair is decodable.

%\begin{figure}
%    \centering
%    \includegraphics[width=3.in]{Evo_plane_P2_type1.eps}
%    \caption{Identity plane, evolution plane, and evolution path for $y^{(1)'}$.}
%    \label{fig:evolution_plane_type1}
%\end{figure}
%
%\begin{figure}
%    \centering
%    \includegraphics[width=3.in]{Evo_plane_P2_type2.eps}
%    \caption{Identity plane, evolution plane, and evolution path for $y^{(2)'}$.}
%    \label{fig:evolution_plane_type2}
%\end{figure}

With the above observation, we propose a new sufficient condition for a degree distribution pair to be decodable: {\it {\color{black} There exists an evolution path that lies entirely beneath the identity plane}}. Hence, the optimization problem becomes to find a degree distribution pair with the evolution path all but touches the identity plane.
%Unlike the conventional single-type IRSA, it is not straightforward to solve this optimization problem via linear programming as we don't know the best evolution path beforehand.
We then adopt the differential evolution technique \cite{DE_book_price05} to solve this optimization problem. Specifically, in the mutation step of differential evolution, we adopt the DEEP algorithm in \cite{DEEP} and use the ``bounce-back" method \cite[Chapter 4.3.1]{DE_book_price05} to handle the boundary conditions. Some optimized degree distributions $\mc{L}\upt[x]$ with $\dmax\upt=8$ for $t\in[T]$ are shown in Table~\ref{tbl:degree_dist}. In addition, the corresponding thresholds $\eta^*$, the analytic result of $\eta$ in the limit as $N\rightarrow\infty$, are also shown. In this table, the policies $\msf{P}_1$, $\msf{P}_2$, and $\msf{P}_3$ are produced by solving the above optimization problem for the cases when $K^{(1)}=K^{(2)}$, $3K^{(1)}=K^{(2)}$, and $7K^{(1)}=K^{(2)}$, respectively.
%Also, the policies $P_4$, $P_5$, and $P_6$ are also produced by the above optimization problem with the stability condition \eqref{eqn:stable_con} for the cases when $K^{(1)}=K^{(2)}$, $K^{(1)}=3K^{(2)}$, and $K^{(1)}=7K^{(2)}$, respectively.
We also provide an example with $T=3$ whose density evolution and the corresponding optimization problem can be found in Appendix~\ref{apx:T_type_DE}. The policy $\msf{P}_4$ provides a set of optimized degree distributions for $T=3$ with $K^{(1)}=K^{(2)}=K^{(3)}$. Finally, the policy $\msf{P}_5$ is that for conventional IRSA with $T=1$ and is directly borrowed from \cite{Liva2011:IRSA_TCOM}. These policies will be tested with simulations in the next section.

\begin{remark}
    From Table~\ref{tbl:degree_dist}, one can observe that every proposed policy has an threshold larger than $100\%$. At first glance, this seems violating the fundamental limit of UMA that at most 1 packet can be successfully delivered in 1 slot, even under perfect coordination. However, we would like to stress that this is not the fundamental limit for our setting, as in our protocol, leveraging heterogeneity among users, it is entirely possible that one can use intra-SIC to decode at most $T$ packets in a slot.
\end{remark}

\begin{remark}
    It is worth mentioning that for $T=2$, when the numbers of active users of type $t\in[2]$ are the same, the optimized degree distributions $\mc{L}\upt[x]$ become the same for $t\in[2]$. See for e.g. $\msf{P}_1$ in Table~\ref{tbl:degree_dist}. This makes perfect sense as when $K^{(1)}= K^{(2)}$, the decodable sets $\mc{D}^{(1)}$ and $\mc{D}^{(2)}$ become symmetric as shown in Example~\ref{ex:decodable_set}. However, when $T=3$, the degree distributions become different. This can also be explained by the different decodable sets in Example~\ref{ex:decodable_set}.
\end{remark}

\begin{table*}
\centering
\caption{Optimized degree distributions and thresholds}
 \label{tbl:degree_dist}
\begin{tabular}{|c|c|l|c|}
\hline
 Policy& $T$ & \multicolumn{1}{c|}{$\mc{L}\upt[x]$} & $\eta^*$ \\ \hline
 $\msf{P}_1$& 2 &
 \begin{tabular}{@{}l@{}}$\mc{L}^{(1)}[x]=0.665x^2+0.1515x^3+0.1835x^8 $ \\ $\mc{L}^{(2)}[x]=0.665x^2+0.1515x^3+0.1835x^8 $\end{tabular}
 &$1.433 $  \\ \hline
 $\msf{P}_2$& 2 &
 \begin{tabular}{@{}l@{}}$\mc{L}^{(1)}[x]=0.3305x^2+0.0165x^4+ 0.0019x^5+0.01825x^6+0.0141x^7+0.4545x^8 $ \\ $\mc{L}^{(2)}[x]=0.4910x^2+0.3145x^3+0.0028x^5+0.0029x^6+0.0287x^7+0.1601x^8 $\end{tabular}
 &$1.232$  \\ \hline
 $\msf{P}_3$& 2 &
 \begin{tabular}{@{}l@{}}$\mc{L}^{(1)}[x]=0.9388x^2+0.0032x^4+0.058x^5 $ \\ $\mc{L}^{(2)}[x]=0.508x^2+0.276x^3+0.216x^8 $\end{tabular}
 &$1.064 $  \\
 \hline
 %$\msf{P}_4$& 2 &
% \begin{tabular}{@{}l@{}}$\mc{L}^{(1)}[x]=0.4985x^2+0.3558x^3+0.1457x^8 $ \\ $\mc{L}^{(2)}[x]=0.4985x^2+0.3558x^3+0.1457x^8$\end{tabular}
% &$1.387 $  \\ \hline
% $\msf{P}_5$& 2 &
% \begin{tabular}{@{}l@{}}$\mc{L}^{(1)}[x]=0.6996x^2+0.2702x^3+0.0302x^6 $ \\ $\mc{L}^{(2)}[x]=0.4245x^2+0.397x^3+0.1785x^8 $\end{tabular}
% &$1.187 $  \\ \hline
% $\msf{P}_6$& 2 &
% \begin{tabular}{@{}l@{}}$\mc{L}^{(1)}[x]=0.864x^2+0.09x^3+0.046x^5 $ \\
%$\mc{L}^{(2)}[x]=0.425x^2+0.337x^3+0.238x^8$
%\end{tabular}
% &$1.043$  \\
% \hline
 $\msf{P}_4$& 3 &
 \begin{tabular}{@{}l@{}l@{}}$\mc{L}^{(1)}[x]=0.746x^2+0.093x^3+0.161x^8 $ \\ $\mc{L}^{(2)}[x]=0.7507x^2+0.0846x^3+0.1647x^8 $ \\
 $\mc{L}^{(3)}[x]=0.7507x^2+0.0846x^3+0.1647x^8 $\end{tabular}
 &$1.851$  \\
 \hline
 $\msf{P}_5$ & 1 & $\mc{L}^{(1)}[x]=0.5x^2+0.28x^3+0.22x^8 $ \cite{Liva2011:IRSA_TCOM}                       &$0.938 $  \\ \hline
\end{tabular}
\end{table*}

%
%Remarks:
%
%$
%P_1:\  K_1 = K_2, \ \ without\ stability\ condition
%$
%
%$
%P_2: 3K_1 = K_2, \ \ without\ stability\ condition
%$
%
%$
%P_3: 7K_1 = K_2, \ \ without\ stability\ condition
%$
%
%$
%P_4:\  K_1 = K_2, \ \ with\ stability\ condition
%$
%
%$
%P_5: 3K_1 = K_2, \ \ with\ stability\ condition
%$
%
%$
%P_6: 7K_1 = K_2, \ \ with\ stability\ condition
%$
%
%$
%P_7:\  K_1 = K_2 = K_3, \ \ three\ types\ heterogeneous
%$
%
%
%$
%P_8: Optimized\  policy\  in\ OMA\ scheme
%$

%%%%%%%%%%%%%%%%%%%%%%%%%%%%%%%%%%%%%%%%%%%%%%%%%%%%%%%%%%%%%%%%%%%%%%%%%%%%%%%%%%%%%%%%%%
\section{Simulation Results}\label{sec:simulation}
\textcolor{black}{In this section, we validate the proposed IRSA with NOMA via extensive simulations under practical frame sizes.} In particular, to demonstrate the effectiveness of the proposed method, we will use the efficiency $\hat{\eta}$ versus the total active users per slot, $\sum_{t=1}^N K\upt/N$ (that is $\eta$), as the metric.
%When comparing the policies obtained by solving the proposed optimization problem with stability condition and that without it, we will also use the packet loss rate,
%\begin{equation}
%    P_e=\frac{\sum_{t=1}^T (K\upt-\hat{K}\upt)}{\sum_{t=1}^T K\upt},
%\end{equation}
%versus the total active users per slot as the metric.

\begin{figure}
    \centering
    \includegraphics[width=3.8in]{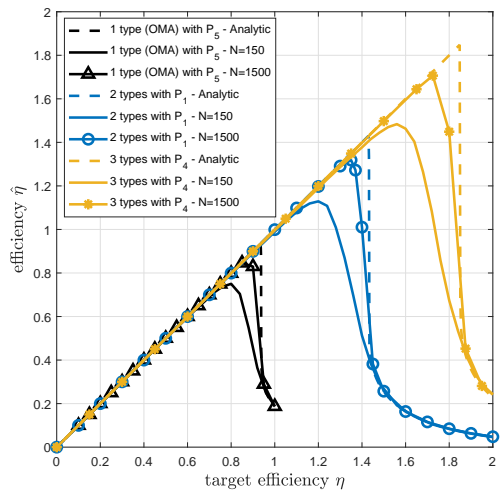}
    \caption{Efficiency versus target efficiency (active users per slot) at $N=150$ and $1500$ for $\msf{P}_5$ ($T=1$), $\msf{P}_1$ ($T=2$), and $\msf{P}_4$ ($T=3$).}
    \label{fig:3types_N_150}
\end{figure}

In Fig.~\ref{fig:3types_N_150}, we compare the achieved efficiency of the proposed IRSA with NOMA for different $T$. Specifically, in Fig.~\ref{fig:3types_N_150}, for both $N=150$ and $1500$, we compare the efficiencies of $\msf{P}_5$, $\msf{P}_1$, and $\msf{P}_4$, that are optimized for $T=1$, $T=2$, and $T=3$, respectively. From these figures, we can observe that the proposed multi-dimensional density evolution indeed accurately predicts the efficiency of the proposed IRSA with NOMA as $N$ increases. Moreover, we observe that the more types, the higher efficiency the proposed algorithm can achieve. The proposed multi-dimensional density evolution predicts an efficiency of $143\%$ for $\msf{P}_1$ and $185\%$ for $\msf{P}_4$, which are both significantly higher than the $93.8\%$ achieved by the conventional IRSA with $\msf{P}_5$. This indicates that when the application at hand presents natural heterogeneity, the proposed IRSA with NOMA can effectively exploit it.

%\begin{figure}
%    \centering
%    \includegraphics[width=3.8in]{3types_N300.eps}
%    \caption{Efficiency versus active users per slot at $N=300$ for $\msf{P}_8$ (T=1), $\msf{P}_1$ ($T=2$), and $\msf{P}_7$ ($T=3$).}
%    \label{fig:3types_N300}
%\end{figure}

\begin{figure}
    \centering
    \includegraphics[width=3.8in]{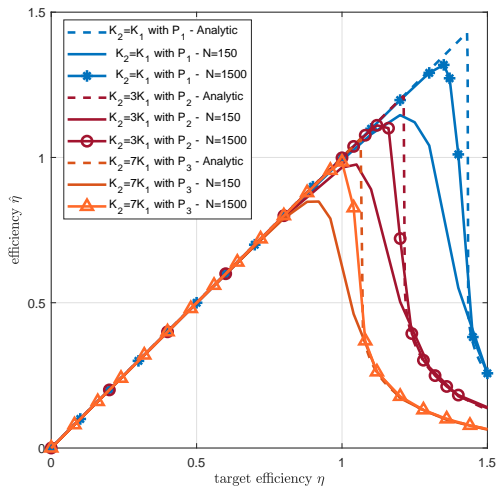}
    \caption{Efficiency versus target efficiency (active users per slot) at $N=150$ and $1500$ for $\msf{P}_1$ ($K^{(1)}=K^{(2)}$), $\msf{P}_2$ ($K^{(1)}=3K^{(2)}$), and $\msf{P}_3$ ($K^{(1)}=7K^{(2)}$).}
    \label{fig:DPDU}
\end{figure}

In Fig.~\ref{fig:DPDU}, we show the efficiency of the proposed IRSA with NOMA for $T=2$ where the two groups have different numbers of users. Again, for both $N=150$ and $1500$, we plot the efficiencies of $\msf{P}_1$, $\msf{P}_2$, and $\msf{P}_3$, that are optimized for the cases $K^{(1)}=K^{(2)}$, $3K^{(1)}=K^{(2)}$, and $7K^{(1)}=K^{(2)}$, respectively. One can see from this figure that the larger the difference between the numbers of users, the smaller efficiency. This is because the small difference means that the number of users in a group is roughly the same from group to group; thereby, it means the heterogeneity is large and more decoding opportunity may be introduced by intra-slot SIC. On the contrary, if the difference is large, it means that one of the group contains most of the users in the network; hence, heterogeneity is small. An extreme example can be seen by considering $M\cdot K^{(1)}= K^{(2)}$ and let $M\rightarrow\infty$, under which the problem would become the homogeneous setting and no heterogeneity can be exploited.

{\color{black}
Figs.~\ref{fig:3types_N_150} and \ref{fig:DPDU} have demonstrated that thanks to NOMA, the proposed protocol can efficiently exploit the heterogeneity inherent in the network. One natural question arising at this point is whether this gain comes solely from the intrinsic benefit of NOMA or the analysis and optimization proposed in Sections \ref{subsec:density_evo} and \ref{subsec:opt_problem} indeed play a role. In other words, do the optimized policies in Table~\ref{tbl:degree_dist} indeed provide non-negligible gains over $\mathsf{P}_5$ when $T=2$ and $T=3$? We note that with a larger decodable set enabled by intra-slot SIC, each node should transmit less in order to maintain the same chance to be successfully decoded while reducing the probability of causing unresolvable collisions. However, $\mathsf{P}_5$ is optimized for $T=1$ \cite{Liva2011:IRSA_TCOM}, which do not take the larger decodable set into account. Hence, nodes adopting this distribution would tend to over-transmit. To confirm this, we look into the degree distributions and observe that the average degrees are 3.6, 3.25, and 3.068 for $\mathsf{P}_5$, $\mathsf{P}_1$, and $\mathsf{P}_4$, respectively. Moreover, for the optimized degree distributions, as $T$ increases, the fraction of degree 2 nodes increases while that of degrees 3 and 8 nodes decreases. In Fig.~\ref{fig:non_opt_150}, we compare the performance of $\mathsf{P}_5$ optimized for $T=1$ and the respective optimized degree distributions at $N=150$. The analytic results obtained by the multi-dimensional density evolution are also plotted, which indicate that the optimized degree distributions provide efficiency gains of 0.083 and 0.174 over $\mathsf{P}_5$ when $T=2$ and $T=3$, respectively. Simulation results also show similar gains, which corroborate our analysis. This gain of the optimized degree distribution over $\mathsf{P}_5$ is expected to increase as $T$ increases due to the larger and larger decodable set.

In contrast, when the numbers of users of different types are drastically different, the gain of the optimized degree distribution over $\mathsf{P}_5$ becomes negligible. For such a scenario, we end up focusing more and more on the type with the largest size and the optimized degree distribution becomes more and more like $\mathsf{P}_5$. This is evident by observing that $\mathsf{P}_5$ and $\Lambda^{(2)}$ (for the type that is 7 times larger than the other) in $\mathsf{P}_3$ are very similar to each other and that the density evolution shows almost identical efficiency. That said, the optimized degree distribution still enjoys a significantly less average degree of 2.89 as opposed to 3.6 in $\mathsf{P}_5$.

\begin{figure}
    \centering
    \includegraphics[width=4in]{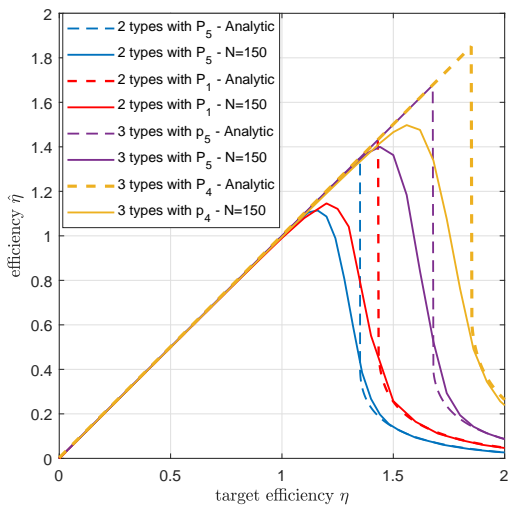}
    \caption{Comparison of the suboptimal policy ($\mathsf{P}_5$) and the optimized policy ($\mathsf{P}_1$ when $T=2$ and $\mathsf{P}_4$ when $T=3$) at $N=150$.}
    \label{fig:non_opt_150}
\end{figure}
}

%Now, to see the influence of the stability condition, we compare the performance of policies $\msf{P}_1$ (without stability) with that of $\msf{P}_4$ (with stability). In Fig.~\ref{fig:stability}, one observes that in terms of efficiency, the stability condition does not make much difference as illustrated in Remark~\ref{rmk:stability}. To see the benefit of including the stability condition, we simulate the packet loss rate $P_e$. We set $N=200$ and repeat the simulation for many frames until at least 500 packet errors are observed. The average packet loss rates are then shown in Fig.~\ref{fig:stability_PLR}, in which we can see that the policy $P_4$ (obtained by solving the optimization problem with stability condition) indeed improves the average packet loss rate.
%\begin{figure}
%    \centering
%    \includegraphics[width=3.8in]{BP_N200.eps}
%    \caption{Efficiency versus active users per slot at $N=150$ and $1500$ for $\msf{P}_1$ (w/o. stability) and $\msf{P}_4$ (w. stability).}
%    \label{fig:stability}
%\end{figure}
%
%\begin{figure}
%    \centering
%    \includegraphics[width=3.8in]{FS_PLR.eps}
%    \caption{Packet loss rate $P_e$ for $\msf{P}_1$ (w/o. stability) and $\msf{P}_4$ (w. stability).}
%    \label{fig:stability_PLR}
%\end{figure}

\section{Extension to Frame Asynchronous Case}\label{sec:FA}
In this section, we extend the proposed IRSA with NOMA and the corresponding multi-dimensional density evolution to the frame asynchronous setting. Both the analysis and the simulation results indicate that similar to \cite{iAmat17}, such asynchrony results in the boundary effect \cite{urbanke11_SC_LDPC} that facilitates our system design \textcolor{black}{by enabling easily implemented regular left degrees to be asymptotically optimal}. Moreover, since there is no concept of frame, the users do not have to wait until the next frame and can immediately start the transmission upon the arrival of a packet.

\subsection{Problem and Protocol}
\begin{figure*}
    \centering
    \includegraphics[width=3.5in]{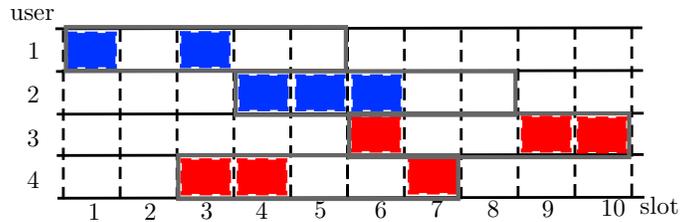}
    \caption{Frame asynchronous ALOHA with two types of heterogeneous users, where users 1 and 2 belong to the type 1 and users 3 and 4 belong to the type 2. Each user has its own view about frame of size $5$ slots, starting from the arrival of its packet. }
    \label{fig:FA_ALOHA_graph}
\end{figure*}

In the frame asynchronous setting, the users are allowed to join the network without frame synchronization; but slots are still synchronous. We assume that the arrival of type $t$ users' data follows a Poisson distribution with arrival rate $g\upt$. That is, the probability that $m\upt$ type $t$ users join the system in a given slot is given by
%\begin{equation}
 $   \frac{\exp(-g\upt)(g\upt)^{m\upt}}{m\upt !}.$
%\end{equation}
We note here that $g\upt$ has the unit ``users per slot" and plays a similar role with the target efficiency $\eta\upt$ in the frame synchronous setting. Also, the sum arrival rate $g=\sum_{t=1}^T g\upt$ plays a similar role with $\eta$ in the frame synchronous setting. Under this frame asynchronous setting, each user has its local view about ``frame" of size $N$ slots starting from the arrival of its packet. For example, for a user joining the network at slot $i$, its local view of frame comprises the slots $[i: i+N-1]$ as this user can transmit its packet in one or multiple of slots within $[i: i+N-1]$. An illustration of this model can be found in Fig.~\ref{fig:FA_ALOHA_graph}.
Under the Poisson arrival process, by the superposition of independent Poisson distributions, the number of active users of type $t$, namely $K\upt$, at slot $i$ again follows a Poisson distribution of rate $\mu\upt_i$, where\footnote{\textcolor{black}{This corresponds to the model with a boundary effect in \cite{iAmat17}. As discussed therein, this boundary effect is possible if the receiver is turned on and monitors users' activities before their transmissions or if we artificially create a guard interval occasionally.}}
\begin{equation}
    \mu\upt_i = \begin{cases}
                  ig\upt, & \mbox{\text{for $1\leq i < N$}} \\
                  Ng\upt, & \mbox{\text{for $i\geq N$}}.
                \end{cases}
\end{equation}
The total number of active users at slot $i$ then follows a Poisson distribution of rate $\mu_i=\sum_{t=1}^T \mu\upt_i$.

In the presence of frame asynchrony, the proposed IRSA with NOMA protocol is modified as follows. Upon the arrival of its packet at slot $i$, a type $t$ user samples from a degree distribution $L\sim\mc{L}\upt$ to determine the number of replicas it sends within its local view of frame. It then immediately sends one packet in slot $i$ and uniformly selects $L-1$ slots from $\Ii=[i+1:i+N-1]$ for sending the remaining $L-1$ replicas. In the decoding process, we again allow both intra-slot and inter-slot SIC, but with a sliding window fashion \cite{sliding_window, iAmat17}. Specifically, to decode the packets of the users joining at slot $i$, the decoder considers a sliding window of size $5N$, namely $[i:i+5N-1]$. The problem can then be treated as a realization of the frame synchronous IRSA with NOMA with frame size $5N$ and can therefore be decoded in a similar fashion.

%\begin{remark}
%    In spatially-coupled LDPC codes, it is shown that by spatially coupling many LDPC codes, a boundary effect created by extra check nodes at around the boundary can help bootstrap the decoding process, which eventually leads to the threshold saturation \cite{urbanke11_SC_LDPC}. Here, similar to \cite{iAmat17}, the same threshold saturation can be expected as the frame asynchronous setting naturally provides a boundary effect.
%\end{remark}

\subsection{Proposed Multi-Dimensional Density Evolution}
To analyze the asymptotic performance of the proposed IRSA with NOMA in the presence of frame asynchrony, we extend the proposed multi-dimensional density evolution to the frame asynchronous setting. Here, for clearly delivering our innovation, we focus on $T=2$ again. Note that the extension to the general $T$ can be completed in the similar way to the synchronous case in Appendix~\ref{apx:T_type_DE}. We recall that the definitions of the node perspective degree distributions $\mc{L}\upt[x], \mc{R}\upt[x]$ and edge perspective degree distributions $\lambda\upt[x], \rho\upt[x]$ are in \eqref{eqn:node_dist}-\eqref{eqn:edge_dist_right}. Moreover, for a type $t$ user active at slot $i$, referred to as a class $i$ type $t$ variable node, its behavior at slot $i$ and that in slots $\Ii$ are different. We thus need to define the node perspective left degree distributions
\begin{align}\label{eqn:node_dist_left_FA}
  \mc{L}\upt\dii[x] = x \quad\text{and}\quad \mc{L}\upt\diIi[x] = \sum_d L\upt_{i\rightarrow \Ii, d} x^d =\sum_d L\upt_d x^{d-1},
\end{align}
where $L\upt_{d,i\rightarrow \Ii}$ is the probability of a type $t$ node joining at slot $i$ that would connect with $d$ of the check nodes in $\Ii$, which is equal to $L\upt_{d+1}$. As for the check nodes corresponding to slot $i$, all the edges must be incident to variable nodes with class $i$ or those with class $j\in\Ki$ where
\begin{equation}
    \Ki  = \begin{cases}
                  \emptyset, & \mbox{\text{for $i=1$}} \\
                  [1:i-1], & \mbox{\text{for $2\leq i <N$}} \\
                  [i-N+1:i-1], & \mbox{\text{for $i \geq N$}}.
                \end{cases}
\end{equation}
We then define the right degree distributions of a class $i$ type $t$ check node that has $d_1$ edges incident to class $i$ type $t$ variable nodes and that of a class $i$ type $t$ check node that has $d_2$ edges incident to type $t$ variable nodes in $\Ki$ as
\begin{align}\label{eqn:node_dist_right_FA}
  \mc{R}\upt\dii[x] = \sum_{d_1} R\upt_{i\rightarrow i, d_1} x^{d_1} \quad\text{and}\quad  \mc{R}\upt\diKi[x] = \sum_{d_2} R\upt_{i\rightarrow \Ki, d_2} x^{d_2},
\end{align}
respectively.

The edge perspective degree distributions can be similarly derived as
\begin{align}\label{eqn:edge_dist_left_FA}
  \lambda\upt\dii[x] = 1 \quad\text{and} \quad \lambda\upt\diIi[x] = \frac{\mc{L}^{'(t)}\diIi[x]}{\mc{L}^{'(t)}\diIi[1]} = \sum_d \lambda\upt_{i\rightarrow \Ii, d} x^{d-2} ,
\end{align}
and
\begin{align}\label{eqn:edge_dist_right_FA}
  \rho\upt\dii[x] = \frac{\mc{R}^{'(t)}\dii[x]}{\mc{R}^{'(t)}\dii[1]}= \sum_{d_1} \rho\upt_{i\rightarrow i, d_1} x^{d_1-1} \quad\text{and}\quad
  \rho\upt\diKi[x] = \frac{\mc{R}^{'(t)}\diKi[x]}{\mc{R}^{'(t)}\diKi[1]}= \sum_{d_2} \rho\upt_{i\rightarrow \Ki, d_2} x^{d_2-1}.
\end{align}
We note that with the proposed protocol, similar to \cite[Prop. 1]{iAmat17}, we have
\begin{proposition}\label{prop:poisson}
    \begin{equation}
        \mc{R}\upt\dii[x]=\rho\upt\dii[x]=\exp(-g\upt(1-x)),
    \end{equation}
    and
    \begin{equation}
        \mc{R}\upt\diKi[x]=\rho\upt\diKi[x]=\exp\left(-\frac{\delta\upt_i(\mc{L}^{'(t)}[1]-1)}{N-1}(1-x)\right),
    \end{equation}
    where $\delta\upt_i=\min(i-1,N-1)g\upt$.
\end{proposition}

For $t\in[T]$, let $x_{\itoi,\ell}\upt$ and $x_{\itoj,\ell}\upt$ be the average erasure probability of the message passed along an edge from a class $i$ type $t$ variable node to a class $i$ type $t$ check node and that of the message passed along an edge from a class $i$ type $t$ variable node to a class $j$ type $t$ check node in iteration $\ell$, respectively. Also, we denote by $y_{\itoi,\ell}\upt$ and $y_{\itoj,\ell}\upt$ the average erasure probability of the message passed along an edge from a class $i$ type $t$ check node to a class $i$ type $t$ variable node and that of the message passed along an edge from a class $i$ type $t$ check node to a class $j$ type $t$ variable node in iteration $\ell$, respectively. In the sequel, we study how these average erasure probabilities evolve with iterations.

The average erasure probability of a message from a type $t$ check node in $\Ii$ to a class $i$ type $t$ variable node in iteration $\ell$ is given by
%\begin{equation}\label{eqn:y_i_avg}
$    \tilde{y}_{i,\ell}\upt = \frac{1}{N-1}\sum_{j\in\Ii} y_{\jtoi,\ell}\upt.$
%\end{equation}
Then, similar to \cite{iAmat17}, we have
\begin{align}
  x_{\itoi,\ell}\upt = \mc{L}\upt[\tilde{y}_{i,\ell}\upt]~\quad\text{and}\quad x_{\itoj,\ell}\upt = y_{\itoi,\ell}\upt \lambda\upt\diIi[\tilde{y}_{i,\ell}\upt].
\end{align}
Also, the average erasure probability of a message from a type $t$ variable node in $\Ki$ to a class $i$ type $t$ check node in iteration $\ell$ is given by
\begin{equation}\label{eqn:x_i_avg}
    \tilde{x}_{i,\ell}\upt = \begin{cases}
                  0, & \mbox{\text{for $i=1$}} \\
                  \frac{1}{i-1}\sum_{k\in\Ki} x_{\ktoi,\ell}\upt, & \mbox{\text{for $2\leq i <N$}} \\
                  \frac{1}{N-1}\sum_{k\in\Ki} x_{\ktoi,\ell}\upt, & \mbox{\text{for $i \geq N$}}.
                \end{cases}
\end{equation}
%An illustration of $\tilde{y}_{i,\ell}\upt$ and $\tilde{x}_{i,\ell}\upt$ can be found in Fig.~\ref{fig:FA_DE_ex}-(a) and Fig.~\ref{fig:FA_DE_ex}-(b), respectively.
%\begin{figure}
%    \centering
%    \includegraphics[width=2.in]{FA_DE_ex.eps}
%    \caption{Illustration of $\tilde{y}_{i,\ell}\upt$ and $\tilde{x}_{i,\ell}\upt$.}
%    \label{fig:FA_DE_ex}
%\end{figure}

With the above results and Proposition~\ref{prop:poisson}, we are now ready to present the multi-dimensional density evolution for the proposed IRSA with NOMA under the frame-asynchronous setting.
\begin{proposition}
    For $t\in[2]$, define
    \begin{align}
        f_1(t) &= \exp(-g\upt x_{\itoi,\ell}\upt), \\ %1
        f_2(t) &= \exp\left(-\frac{\delta_i\upt(\mc{L}^{'(t)}[1]-1)}{N-1}\tilde{x}_{i,\ell}\upt \right), \\ %2
        f_3(t) &= x_{\itoi,\ell}\upt g\upt \exp(-g\upt x_{\itoi,\ell}\upt), \\ %5
        f_4(t) &= \tilde{x}_{i,\ell}\upt \frac{\delta_i\upt(\mc{L}^{'(t)}[1]-1)}{N-1}\exp\left(-\frac{\delta_i\upt(\mc{L}^{'(t)}[1]-1)}{N-1}\tilde{x}_{i,\ell}\upt \right). %6
    \end{align}
    The multi-dimensional density evolution is as follows. For $t\in[2]$, $i\in\mbb{N}$, and $j\in\Ki$, we have
    \begin{align}\label{eqn:DE_FA}
        y_{\itoi,\ell+1}\upt=y_{\itoj,\ell+1}\upt = 1-f_1(t)f_2(t) \prod_{\bar{t}\in [2],\bar{t}\neq t}\left(f_1(\bar{t})f_2(\bar{t})+f_3(\bar{t})f_2(\bar{t})+ f_1(\bar{t})f_4(\bar{t})\right).
    \end{align}
\end{proposition}
The proof of this result is omitted for the sake of brevity. But we point out that in \eqref{eqn:DE_FA}, for $\bar{t}\neq t$, $(f_1(\bar{t})f_2(\bar{t})+f_3(\bar{t})f_2(\bar{t})+ f_1(\bar{t})f_4(\bar{t}))$ corresponds to the average probability that there is at most one erasure among edges connecting to a type $\bar{t}$ check node.

\subsection{Simulation Results}
\begin{figure}
    \centering
    \includegraphics[width=3.5in]{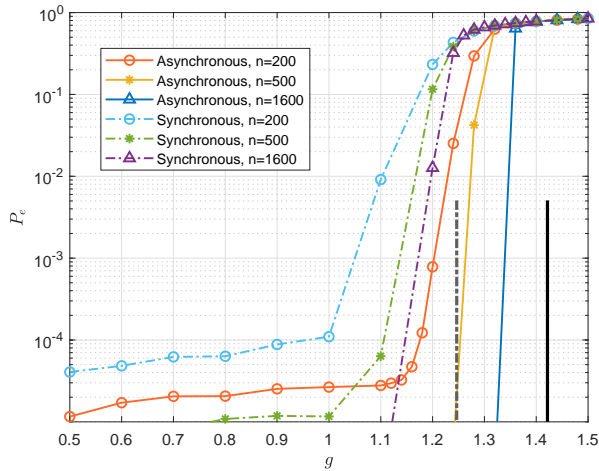}
    \caption{Packet loss rate versus sum arrival rate $g$ in the frame asynchronous setting. The thresholds obtained by the proposed density evolution are also plotted.}
    \label{fig:FA-CSA-compare}
\end{figure}

\textcolor{black}{First, the threshold of the proposed IRSA with NOMA for the left-regular $\Lambda\upt=x^3$ is evaluated under the frame asynchronous setting as $g=1.42$. Note that for the same left-regular degree distribution, the density evolution shows a $g=1.24$ threshold for the frame synchronous setting. \textcolor{black}{This shows that, as predicted, the boundary effect helps bootstrapping the decoding process and allows the left-regular degree distribution $x^3$ to achieve a threshold that is close to $g=1.433$ achieved by the optimized degree distribution $\mathsf{P}_1$ in Table~\ref{tbl:degree_dist} in the frame synchronous setting.} Simulation results are then provided in Fig.~\ref{fig:FA-CSA-compare}, where the packet loss rate $P_e$ versus sum arrival rate $g$ is plotted for $N=200$, $500$, and $1600$.}

\section{Conclusion}\label{sec:conclude}
In this paper, we have investigated UMA in the presence of heterogeneous users. A novel protocol, IRSA with NOMA, has been proposed to leverage the heterogeneity inherent in the problem. To analyze the proposed protocol, a novel multi-dimensional density evolution has been proposed, which has been shown to be able to accurately predict the asymptotic performance of IRSA with NOMA under the modified peeling decoding. An optimization problem has then been formulated and solved for finding optimal degree distributions for our IRSA with NOMA. Simulation results have demonstrated that the proposed IRSA with NOMA can exploit the natural heterogeneity and obtained efficiency higher than that achieved by conventional IRSA. \textcolor{black}{Finally, an extension of the proposed protocol to the frame-asynchronous setting has been investigated, where a boundary effect that bootstraps decoding process has been discovered via both analysis and simulation.} %In the second extension, a means to create heterogeneity among homogeneous users has been introduced, so that UMA with homogeneous users can also benefit from the proposed IRSA with NOMA.

\textcolor{black}{Throughout the paper, we have taken the MAC layer perspective. One potential future work is to extend our idea to a more practical model as in \cite{Gaudenzi17power_unbalance} and to see how randomly selecting power can further improve the performance in the presence of heterogeneity inherent in the network. Another interesting direction is to analyze how much more we can gain from having more types. From Fig.~\ref{fig:3types_N_150}, we have already observed diminishing returns going from $T=2$ to $T=3$. We would expect that the $\eta^*$ converges to a constant as $T\rightarrow \infty$. The analysis of this convergence is left for future work.}

%\section*{Acknowledgment}
%The authors would like to thank Prof. Krishna R. Narayanan at Texas A\&M University for helpful discussions.

%\section*{Acknowledgment}
%The authors would like to thank Mr. Shih-Hung Yang at National Chiao Tung University for assisting the simulations.

\bibliographystyle{IEEEtran}
\bibliography{journal_abbr,bib_5g}

\newpage
\clearpage\pagenumbering{arabic}
\appendices
\section{Multi-Dimensional Density Evolution for General $T$}\label{apx:T_type_DE}
In this appendix, we discuss the multi-dimensional density evolution for general $T$.

\subsection{Density Evolution for General $T$}
For $t\in[T]$, we recall that $x_\ell\upt$ and $y_\ell\upt$ are the average erasure probability of the message passed along an edge from $\Vt$ to $\Ct$ and that of the message passed along an edge from $\Ct$ to $\Vt$ in iteration $\ell$, respectively.
%To initialize at $\ell=0$, we observe that for an edge connected to a type $t$ check node $\Ct$, the message is revealed (not in erasure) if and only if a) $\Ct$ has degree 1; and b)  $\msf{c}^{(\bar{t})}$ has a degree either 0 or 1 for every $\bar{t}\in[T]\setminus\{t\}$. Hence, the average erasure probability is given by
%\begin{equation}
%    y_0\upt = 1-\rho\upt[0]\cdot\prod_{\bar{t}=1,\bar{t}\neq t}^T \left( \mc{R}^{(\bar{t})}[0] + \mc{R}^{'(\bar{t})}[0] \right),
%\end{equation}
%where $\mc{R}^{(\bar{t})}[0]=R^{(\bar{t})}_0$ and $\mc{R}^{'(\bar{t})}[0]=R^{(\bar{t})}_1$ are the fractions of type $\bar{t}$ check nodes having degree 0 and degree 1, respectively.
In iteration $\ell$, for an edge incident to a variable node $\Vt$ with degree $d$, the only possibility that the message along this edge to a $\Ct$ is in erasure is that all the other $d-1$ edges are in erasure. Therefore, the probability that the message passed along this edge is in erasure is $(y_\ell\upt)^{d-1}$. Now, averaging over all the edges results in the average erasure probability
\begin{equation}\label{eqn:DE_x_T}
    x_\ell\upt = \sum_d \lambda\upt_d (y_\ell\upt)^{d-1} = \lambda\upt [y_\ell\upt ].
\end{equation}
\textcolor{black}{We denote by $\mathbf{c}$ the vector whose $t$-th entry stores the number of unresolved (erased) packets. For an edge incident to a check node $\Ct$ with degree $d_t$, the message passed along this edge to a $\Vt$ is revealed if and only if the corresponding $\mathbf{c}$ belongs to the decodable set $\mc{D}\upt$.
%a) all the other $d-1$ edges incident to $\Ct$ are revealed; and b) all but at most one of the edges incident to $\msf{c}^{(\bar{t})}$ are revealed for every $\bar{t}\in[T]\setminus\{t\}$.
Suppose in the same super check node $n$, $\msf{c}^{(\bar{t})}_n$ has degree $d_{\bar{t}}$. Then the above event has probability
\begin{align}
    &\sum_{\mathbf{c}\in\mc{D}\upt} (1-x_\ell\upt)^{d_t-c_t} \prod_{\bar{t}=1, \bar{t}\neq t}^T \binom{d_{\bar{t}}}{c_{\bar{t}}} (x_\ell^{(\bar{t})})^{c_{\bar{t}}}(1-x_\ell^{(\bar{t})})^{d_{\bar{t}}-c_{\bar{t}}} \nonumber \\
    &=(1-x_\ell\upt)^{d_t-1} \sum_{\mathbf{c}\in\mc{D}\upt} \prod_{\bar{t}=1, \bar{t}\neq t}^T \binom{d_{\bar{t}}}{c_{\bar{t}}} (x_\ell^{(\bar{t})})^{c_{\bar{t}}}(1-x_\ell^{(\bar{t})})^{d_{\bar{t}}-c_{\bar{t}}},
%     \left( (1-x_\ell^{(\bar{t})})^{d_{\bar{t}}} + d_{\bar{t}} x_\ell^{(\bar{t})}(1-x_\ell^{(\bar{t})})^{d_{\bar{t}}-1}\right),
\end{align}
where the equality is due to the fact that every vector in $\mc{D}\upt$ has $c_t=1$.}
%\begin{equation}
%    (1-x_\ell\upt)^{d-1}\cdot\prod_{\bar{t}=1, \bar{t}\neq t}^T \left( (1-x_\ell^{(\bar{t})})^{d_{\bar{t}}} + d_{\bar{t}} x_\ell^{(\bar{t})}(1-x_\ell^{(\bar{t})})^{d_{\bar{t}}-1}\right),
%\end{equation}
%where $(1-x_\ell^{(\bar{t})})^{d_{\bar{t}}}$ is the probability that all $d_{\bar{t}}$ edges are revealed and $d_{\bar{t}} x_\ell^{(\bar{t})}(1-x_\ell^{(\bar{t})})^{d_{\bar{t}}-1}$ is the probability that all but one edges are revealed.
Averaging over all the edges and over all the type $\bar{t}$ check nodes for $\bar{t}\in[T]\setminus\{t\}$ shows that the average probability of correct decoding is given by
\begin{align}\label{eqn:DE_y_intermediate_T}
    &\sum_{d_t} \lambda\upt_{d_t}(1-x_\ell\upt)^{d_t-1}\cdot\nonumber \\
    &\sum_{\mathbf{c}\in\mc{D}\upt}\prod_{\bar{t}=1, \bar{t}\neq t}^T \sum_{d_{\bar{t}}}R^{(\bar{t})}_{d_{\bar{t}}}\binom{d_{\bar{t}}}{c_{\bar{t}}} (x_\ell^{(\bar{t})})^{c_{\bar{t}}}(1-x_\ell^{(\bar{t})})^{d_{\bar{t}}-c_{\bar{t}}} \nonumber \\
    &=\rho\upt[1-x_\ell\upt]\sum_{\mathbf{c}\in\mc{D}\upt}\prod_{\bar{t}=1, \bar{t}\neq t}^T  (x_\ell^{(\bar{t})})^{c_{\bar{t}}}(\mc{R}^{(\bar{t})})^{\{c_{\bar{t}}\}}[1-x_\ell^{(\bar{t})}]/c_{\bar{t}}!.
\end{align}
Therefore, the average erasure probability becomes
\textcolor{black}{\begin{align}\label{eqn:DE_y_T}
    &y_{\ell+1}\upt = 1-\rho\upt[1-x_\ell\upt]\cdot \nonumber \\
    &\sum_{\mathbf{c}\in\mc{D}\upt}\prod_{\bar{t}=1, \bar{t}\neq t}^T  (x_\ell^{(\bar{t})})^{c_{\bar{t}}}(\mc{R}^{(\bar{t})})^{\{c_{\bar{t}}\}}[1-x_\ell^{(\bar{t})}]/c_{\bar{t}}!.
\end{align}}
Plugging \eqref{eqn:DE_x_T} into \eqref{eqn:DE_y_T} leads to the evolution of average erasure probability of a type $t$ check node as shown in \eqref{eqn:DE_full_T} in the bottom of this page.
\begin{figure*}[!b]
\normalsize
\hrulefill
\textcolor{black}{\begin{IEEEeqnarray}{rCl}\label{eqn:DE_full_T}
     y_{\ell+1}\upt=1-\rho\upt[1-\lambda\upt[y_\ell\upt]]\cdot \sum_{\mathbf{c}\in\mc{D}\upt}\prod_{\bar{t}=1, \bar{t}\neq t}^T  \frac{(\lambda^{(\bar{t})}[y_\ell^{(\bar{t})}])^{c_{\bar{t}}}}{c_{\bar{t}}!}(\mc{R}^{(\bar{t})})^{\{c_{\bar{t}}\}}[1-\lambda^{(\bar{t})}[y_\ell^{(\bar{t})}]],\quad t\in[T].
\end{IEEEeqnarray}}
%\setcounter{equation}{\value{tempeqcounter}} % restore correct value
%\vspace*{4pt}
\end{figure*}

\subsection{Convergence and Stability Condition}
After obtaining the density evolution in \eqref{eqn:DE_full_T}, for any given degree distributions $\mc{L}\upt$ (or $\lambda\upt$) and $\mc{R}\upt$ (or $\rho\upt$), we again can analyze whether the average erasure probability converges to 0 by checking whether $y_\ell\upt> y_{\ell+1}\upt$ for every $\ell$ and every $y_\ell\upt>0$.

In what follows, we again derive the stability condition. Similar to the $T=2$ case, we enforce $\lambda_1\upt=0$ for all $t\in[T]$. Assuming $y\upt$ is very small for all $t\in[T]$, we expand the degree distributions and approximate them by keeping only the linear terms as follows,
\textcolor{black}{\begin{align}
    \lambda\upt[y\upt] &\approx \lambda_2\upt y\upt, \label{eqn:stable_1_T} \\
    \rho\upt[1-\lambda\upt[y\upt]] &\approx 1-\rho^{'{(t)}}[1]\lambda_2\upt y\upt, \label{eqn:stable_2_T} \\
    R^{(\bar{t})}[1-\lambda^{(\bar{t})}[y^{(\bar{t})}]] &\approx 1-R^{'(\bar{t})}[1] \lambda_2^{(\bar{t})} y^{(\bar{t})}, \label{eqn:stable_3_T} \\
    (R^{(\bar{t})})^{\{k\}}[1-\lambda^{(\bar{t})}[y^{(\bar{t})}]] &\approx (R^{(\bar{t})})^{\{k\}}[1] \nonumber\\
    &- (R^{(\bar{t})}[1])^{\{k+1\}}\lambda_2^{(\bar{t})}y^{(\bar{t})}.\label{eqn:stable_4_T}
\end{align}}
We can now linearize the recursion around 0 by plugging \eqref{eqn:stable_1_T}-\eqref{eqn:stable_4_T} into \eqref{eqn:DE_full_T} to get the same stability condition
\begin{equation}\label{eqn:stable_con_T}
    \lambda_2\upt < \frac{1}{\rho^{'{(t)}}[1]}\quad\text{for $t\in[T]$}.
\end{equation}

\subsection{Optimization Problem and Optimized Degree Distributions}
Similar to the $T=2$ case, we again apply the Poisson approximation and rewrite the convergence condition in \eqref{eqn:DE_full_T} as \eqref{eqn:DE_full_poisson_T} in the bottom of the next page.
\begin{figure*}[!b]
\normalsize
\hrulefill
\textcolor{black}{\begin{IEEEeqnarray}{rCl}\label{eqn:DE_full_poisson_T}
     y\upt&>&1-\exp\left( -\eta\upt \mc{L}^{'(t)}[1] \lambda\upt[y\upt]\right) \cdot \nonumber \\
     && \sum_{\mathbf{c}\in\mc{D}\upt}\prod_{\bar{t}=1, \bar{t}\neq t}^T \frac{(\lambda^{(\bar{t})}[y_\ell^{(\bar{t})}]\cdot \eta^{(\bar{t})} \mc{L}^{'(\bar{t})}[1])^{c_{\bar{t}}}}{c_{\bar{t}}!}  \exp\left( -\eta^{(\bar{t})} \mc{L}^{'(\bar{t})}[1] \lambda^{(\bar{t})} [y^{(\bar{t})}] \right),\quad t\in[T].
\end{IEEEeqnarray}}
%\setcounter{equation}{\value{tempeqcounter}} % restore correct value
%\vspace*{4pt}
\end{figure*}

%\begin{figure*}[!b]
%\normalsize
%%\setcounter{tempeqcounter}{\value{equation}} % temp store of current value
%\hrulefill
%\begin{IEEEeqnarray}{rCl}\label{eqn:DE_full_poisson}
%%\setcounter{equation}{\value{storeeqcounter}} % number of this equation
%     y_{\ell+1}\upt&=&1-\exp\left( -\eta\upt \mc{L}^{'(t)}[1] \lambda\upt[y_\ell\upt]\right) \cdot \nonumber \\
%     && \prod_{\bar{t}=1, \bar{t}\neq t}^T \left( \exp\left( -\eta^{(\bar{t})} \mc{L}^{'(\bar{t})}[1] \lambda^{(\bar{t})} [y_\ell^{(\bar{t})}] \right)+  \eta^{(\bar{t})} \mc{L}^{'(\bar{t})}[1] \lambda^{(\bar{t})} [y_\ell^{(\bar{t})}] \exp\left( -\eta^{(\bar{t})} \mc{L}^{'(\bar{t})}[1] \lambda^{(\bar{t})} [y_\ell^{(\bar{t})}] \right)\right).
%\end{IEEEeqnarray}
%%\setcounter{equation}{\value{tempeqcounter}} % restore correct value
%%\vspace*{4pt}
%\end{figure*}

Finally, we are able to formulate the optimization problem that maximizes the target efficiency subject to derived conditions:
\begin{subequations}
\begin{alignat*}{2}
&\!\max_{\{\mc{L}\upt[x]\} }        &\qquad& \eta =\sum_{t=1}^T \eta\upt \\
&\text{subject to} &      & \text{convergence condition~}\eqref{eqn:DE_full_poisson_T},\\
%&                  &      & \text{stability condition~}\eqref{eqn:stable_con_T},\\
&                  &      & \lambda_1\upt=0~\text{for all $t\in[T]$},\\
&                  &      & L_d\upt\geq 0, ~\text{for all $d\in[\dmax\upt]$ and $t\in[T]$}, \\
&                  &      & \mc{L}\upt[1]=1~\text{for all $t\in[T]$}.
\end{alignat*}
\end{subequations}
%where $\dmax\upt$ is the maximum degree of $\mc{L}\upt[x]$ that has to be imposed in practice. This problem can then be solved by linear programming \cite{Luby01peeling} or by differential evolution \cite{storn97}.
Similar to $T=2$, one may include the stability condition \eqref{eqn:stable_con_T} into the above optimization problem to make sure that the packet loss rate indeed vanishes as $N\rightarrow \infty$.

%\section{Proof of Lemma \ref{lma:NOMA_power}}\label{apx:lemma_proof}
%We note that for users selecting different power levels, they also use different interleavers. Hence, the codes are independent and the following Gaussian MAC capacity region gives the largest achievable rates \cite{cover91} (assuming the decoding order $1\rightarrow 2 \rightarrow \ldots \rightarrow T$ and unit noise variance without loss of generality)
%\begin{align}\label{eqn:MAC_region}
%  R_t &< \frac{1}{2} \log_2\left(1+\frac{P_t}{1 + \sum_{i=t+1}^{T} P_i}\right), \quad\text{for $t\in[2:T]$,} \nonumber \\
%  R_T &< \frac{1}{2} \log_2\left(1+P_T\right).
%\end{align}
%Clearly, the lemma holds for $j=1$. Moreover, for $j=2$, setting $R_{T-1}=R_{T}$ results in
%\begin{align}
%  \frac{P_{T-1}}{1 + P_T} = P_T ~\Leftrightarrow~ P_{T-1} = P_T + P_{T}^2 =\mc{O}(P_T^2).
%\end{align}
%By induction, assume $P_{T-j+1}=\mc{O}(P_T^j)$ for every $j\in[k-1]$. Now, consider $j=k$. Set $R_{T-k+1}=R_T$ shows that
%\begin{align}
%   \frac{P_{T-k+1}}{1 + \sum_{i=T-k+2}^{T} P_i}= P_T ~\Leftrightarrow~ P_{T-k+1} = \mc{O}(P_T\cdot P_{T-k+2}) =\mc{O}(P_T^{k}).
%\end{align}

\end{document}